\documentclass[a4paper,aps,nofootinbib,floatfix,showpacs,preprintnumbers,
twocolumn,
superscriptaddress]{revtex4-1} 
\usepackage{graphicx}
\usepackage{bm}
\usepackage{mathrsfs}
\usepackage{float}
\usepackage{tabularx} 
\usepackage{amsmath}
\usepackage{epstopdf}
\usepackage{hyperref}
\usepackage{natbib}

\input epsf

\usepackage{amsthm}
\usepackage{mathtools}
\usepackage{physics}
\usepackage{graphicx}
\usepackage[left=23mm,right=13mm,top=35mm,columnsep=15pt]{geometry} 
\usepackage{adjustbox}
\usepackage{placeins}
\usepackage[T1]{fontenc}
\usepackage{lipsum}
\usepackage{csquotes}

\begin{document}

\title{Tagging and localisation of ionizing events using NbSi transition edge phonon sensors for Dark Matter searches}

\author{C. Augier}
\affiliation{Univ Lyon, Universit\'e Lyon 1, CNRS/IN2P3, IP2I-Lyon, F-69622, Villeurbanne, France}
\author{A.~Beno\^{i}t}
\affiliation{Institut N\'{e}el, CNRS/UJF, 25 rue des Martyrs, BP 166, 38042 Grenoble, France}
\author{L.~Berg\'{e}}
\affiliation{Universit\'{e} Paris-Saclay, CNRS/IN2P3, IJCLab, 91405 Orsay, France}
\author{J.~Billard}
\affiliation{Univ Lyon, Universit\'e Lyon 1, CNRS/IN2P3, IP2I-Lyon, F-69622, Villeurbanne, France}
\author{A.~Broniatowski}
\affiliation{Universit\'{e} Paris-Saclay, CNRS/IN2P3, IJCLab, 91405 Orsay, France}
\author{P.~Camus}
\affiliation{Institut N\'{e}el, CNRS/UJF, 25 rue des Martyrs, BP 166, 38042 Grenoble, France}
\author{A.~Cazes}
\affiliation{Univ Lyon, Universit\'e Lyon 1, CNRS/IN2P3, IP2I-Lyon, F-69622, Villeurbanne, France}
\author{M.~Chapellier}
\affiliation{Universit\'{e} Paris-Saclay, CNRS/IN2P3, IJCLab, 91405 Orsay, France}
\author{F.~Charlieux}
\affiliation{Univ Lyon, Universit\'e Lyon 1, CNRS/IN2P3, IP2I-Lyon, F-69622, Villeurbanne, France}
\author{J.~Colas}
\affiliation{Univ Lyon, Universit\'e Lyon 1, CNRS/IN2P3, IP2I-Lyon, F-69622, Villeurbanne, France}
\author{M. De~J\'{e}sus}
\affiliation{Univ Lyon, Universit\'e Lyon 1, CNRS/IN2P3, IP2I-Lyon, F-69622, Villeurbanne, France}
\author{L.~Dumoulin}
\affiliation{Universit\'{e} Paris-Saclay, CNRS/IN2P3, IJCLab, 91405 Orsay, France}
\author{K.~Eitel}
\affiliation{Karlsruher Institut f\"{u}r Technologie, Institut f\"{u}r Astroteilchenphysik, Postfach 3640, 76021 Karlsruhe, Germany}
\author{J.-B.~Filippini}
\affiliation{Univ Lyon, Universit\'e Lyon 1, CNRS/IN2P3, IP2I-Lyon, F-69622, Villeurbanne, France}
\author{D.~Filosofov}
\affiliation{JINR, Laboratory of Nuclear Problems, Joliot-Curie 6, 141980 Dubna, Moscow Region, Russian Federation}
\author{J.~Gascon}\email{j.gascon@ipnl.in2p3.fr}
\affiliation{Univ Lyon, Universit\'e Lyon 1, CNRS/IN2P3, IP2I-Lyon, F-69622, Villeurbanne, France}
\author{A.~Giuliani}
\affiliation{Universit\'{e} Paris-Saclay, CNRS/IN2P3, IJCLab, 91405 Orsay, France}
\author{M.~Gros}
\affiliation{IRFU, CEA, Universit\'{e} Paris-Saclay, F-91191 Gif-sur-Yvette, France}
\author{E.~Guy}
\affiliation{Univ Lyon, Universit\'e Lyon 1, CNRS/IN2P3, IP2I-Lyon, F-69622, Villeurbanne, France}
\author{Y.~Jin}
\affiliation{C2N, CNRS, Universit\'e  Paris-Sud, Universit\'e  Paris-Saclay, 91120 Palaiseau, France}
\author{A.~Juillard}
\affiliation{Univ Lyon, Universit\'e Lyon 1, CNRS/IN2P3, IP2I-Lyon, F-69622, Villeurbanne, France}
\author{H.~Lattaud}
\affiliation{Univ Lyon, Universit\'e Lyon 1, CNRS/IN2P3, IP2I-Lyon, F-69622, Villeurbanne, France}
\author{S.~Marnieros}
\affiliation{Universit\'{e} Paris-Saclay, CNRS/IN2P3, IJCLab, 91405 Orsay, France}
\author{N.~Martini}
\affiliation{Univ Lyon, Universit\'e Lyon 1, CNRS/IN2P3, IP2I-Lyon, F-69622, Villeurbanne, France}
\author{X.-F.~Navick}
\affiliation{IRFU, CEA, Universit\'{e} Paris-Saclay, F-91191 Gif-sur-Yvette, France}
\author{C.~Nones}
\affiliation{IRFU, CEA, Universit\'{e} Paris-Saclay, F-91191 Gif-sur-Yvette, France}
\author{E.~Olivieri}
\affiliation{Universit\'{e} Paris-Saclay, CNRS/IN2P3, IJCLab, 91405 Orsay, France}
\author{C.~Oriol}
\affiliation{Universit\'{e} Paris-Saclay, CNRS/IN2P3, IJCLab, 91405 Orsay, France}
\author{P.~Pari}
\affiliation{IRAMIS, CEA, Universit\'{e} Paris-Saclay, F-91191 Gif-sur-Yvette, France}
\author{B.~Paul}
\affiliation{IRFU, CEA, Universit\'{e} Paris-Saclay, F-91191 Gif-sur-Yvette, France}
\author{D.~Poda}
\affiliation{Universit\'{e} Paris-Saclay, CNRS/IN2P3, IJCLab, 91405 Orsay, France}
\author{A. Rojas}
\affiliation{Univ. Grenoble Alpes, CNRS, Grenoble INP, LPSC/LSM-IN2P3, 73500 Modane, France}
\author{S.~Rozov}
\affiliation{JINR, Laboratory of Nuclear Problems, Joliot-Curie 6, 141980 Dubna, Moscow Region, Russian Federation}
\author{V.~Sanglard}
\affiliation{Univ Lyon, Universit\'e Lyon 1, CNRS/IN2P3, IP2I-Lyon, F-69622, Villeurbanne, France}
\author{L.~Vagneron}
\affiliation{Univ Lyon, Universit\'e Lyon 1, CNRS/IN2P3, IP2I-Lyon, F-69622, Villeurbanne, France}
\author{E.~Yakushev}
\affiliation{JINR, Laboratory of Nuclear Problems, Joliot-Curie 6, 141980 Dubna, Moscow Region, Russian Federation}
\author{A.~Zolotarova}
\affiliation{IRFU, CEA, Universit\'e Paris-Saclay, Saclay, F-91191 Gif-sur-Yvette, France}
\collaboration{EDELWEISS Collaboration} 

\author{B. J. Kavanagh}
\affiliation{Instituto de F\'isica de Cantabria (IFCA, UC-CSIC), Avenida de Los Castros s/n, 39005 Santander, Spain}

\date{\today} 

\begin{abstract}
In the context of direct searches of sub-GeV Dark Matter particles with germanium detectors, the EDELWEISS collaboration has tested a new technique to tag ionizing events using NbSi transition edge athermal phonon sensors.
The emission of the athermal phonons generated by the Neganov-Trofimov-Luke effect associated with the drift of electrons and holes through the detectors is used to tag ionization events generated in  specific parts of the detector localized in front of the NbSi sensor and to reject by more than a factor 5 (at 90\% C.L.) the background from heat-only events that dominates the spectrum above 3 keV.
This method is able to improve by a factor 2.8 the previous limit on spin-independent interactions of 1 GeV/c$^2$ WIMPs obtained with the same detector and data set but without this tagging technique. 

\end{abstract}


\maketitle

\section{Introduction}

The absence of the discovery of a Weakly Interacting Massive Particle (WIMP)~\cite{rev,rev-bertone} by large double-phase detectors~\cite{xenon,xenon-lux,xenon-panda}, coupled to the absence of the observation of beyond-the-Standard-Model particles at the LHC~\cite{lhc}, has revived a keen interest for models where the Dark Matter (DM) particle would have so far escaped detection by having a mass below the GeV/c$^2$ range~\cite{lowmass} and a phenomenology departing from the standard WIMP scenario. 
The scattering of such particles --- present in our galactic halo --- on nuclei
could produce nuclear recoils with kinetic energies well below 1 keV. 

In that context, direct DM search experiments are now developing  techniques to reduce the threshold for the 
detection of recoils (either a nucleus or an electron) to energies in the eV range~\cite{cresst,SuperCDMS-HVeV,CPD,damic,sensei,damic-m}.
The exploration of this new energy domain reveals new background populations that are not fully understood yet~\cite{excess}. 
In particular, experiments using cryogenic detectors have all observed low-energy populations that are not associated with ionization signals, labeled here Heat-Only events (HO). They may be associated with sudden stress release in the detectors~\cite{cresst-crack}. 
In EDELWEISS, these constitute by far the most important background at energies below a few keV~\cite{edwiii,red30,nbsi-migdal}.
At the low energies that are relevant for sub-GeV DM searches, the resolution on the ionization signal of the current EDELWEISS detectors is not sufficient to remove this background.
For this reason, the EDELWEISS collaboration has investigated whether the use of an athermal phonon sensor --- a NbSi Transition Edge Sensor (TES)~\cite{nbsi-ltd}, instead of its usual Ge Neutron Transmutation-Doped (NTD) thermal phonon sensor~\cite{edwtech} --- helps reducing this background. Up to now this approach had limited success~\cite{nbsi-migdal}.
However, this work did not fully exploit the potential of the athermal sensor used therein.

The sensitivity of TES to out-of-equilibrium  phonons has already been reported on several other TES-instrumented bolometers~\cite{cresst-tes,cdms-tes} that have been highly optimized for energy resolution and signal-to-noise ratio.
The position-dependence of the response to athermal phonons of the W-TES sensor array of the SuperCDMS detectors has been used to identify interactions occurring near the sensors~\cite{cdms-phonontime}.
However, this discrimination comes from the presence of high-field regions in front of the sensor as opposed to the low field present in the bulk of the detector.
These are absent in the EDELWEISS NbSi detector due to its quasi-planar electrode design.
However, we will see that the event selection in Ref.~\cite{nbsi-migdal} privileged the sensitivity to ballistic phonons, whose long mean free paths completely wash out any position-dependent effect. This favors energy resolution but fails to remove all sources of Heat-Only events. 
Here, instead, we will exploit the potential of the NbSi film to detect higher-energy phonons in the THz range, with mean free paths significantly shorter than the cm-scale of the detector and emitted in the earliest stage of either the primary interaction or the Neganov-Trofimov-Luke (NTL) process~\cite{Neganov,Luke}. The NTL effect corresponds to the dissipation of the energy imparted to the electrons and holes as they are forced to drift across the detector.

In this paper, we demonstrate the sensitivity 
of the NbSi TES sensor to primary NTL phonons and describe how it can be used to tag ionizing events occurring in a well-defined region of the detector, and thus reject HO events. 
This paper will first present the detector and the experimental setup, and then describe the overall detector performance. 
The following section describes the ionization signal cut used to separate events according to their position inside the detector volume.
This selection is then used to measure the position dependence of the pulse shape of the phonon signal, and to identify a fast component associated to the emission of out-of-equilibrium NTL phonons in the region where the electric field lines reach the NbSi film.
In the final sections, it is shown how this extra signal can be used to identify ionizing events and quantify the rejection of HO events. 
This is further illustrated by comparing the performance of this phonon-based charge tag to the electrode-based tag used with the same detector in the DM search of Ref.~\cite{nbsi-migdal}.

\section{Experimental setup and data set}

The present data were collected in the same cool-down as described in Ref.~\cite{nbsi-migdal}.
Namely, the experimental site is located in the Laboratoire Souterrain de Modane (LSM, France). The detector was operated in the  ultra-low background cryostat of the EDELWEISS collaboration~\cite{edwtech}. 
The detectors are thus protected by a 4800 m.w.e. rock overburden, as well as layers of polyethylene and lead shields described in Ref.~\cite{edwtech}.

\begin{figure}[htp]
\begin{center}
\includegraphics[width=0.90\linewidth,trim= 0in 0.0in 0in 0in]{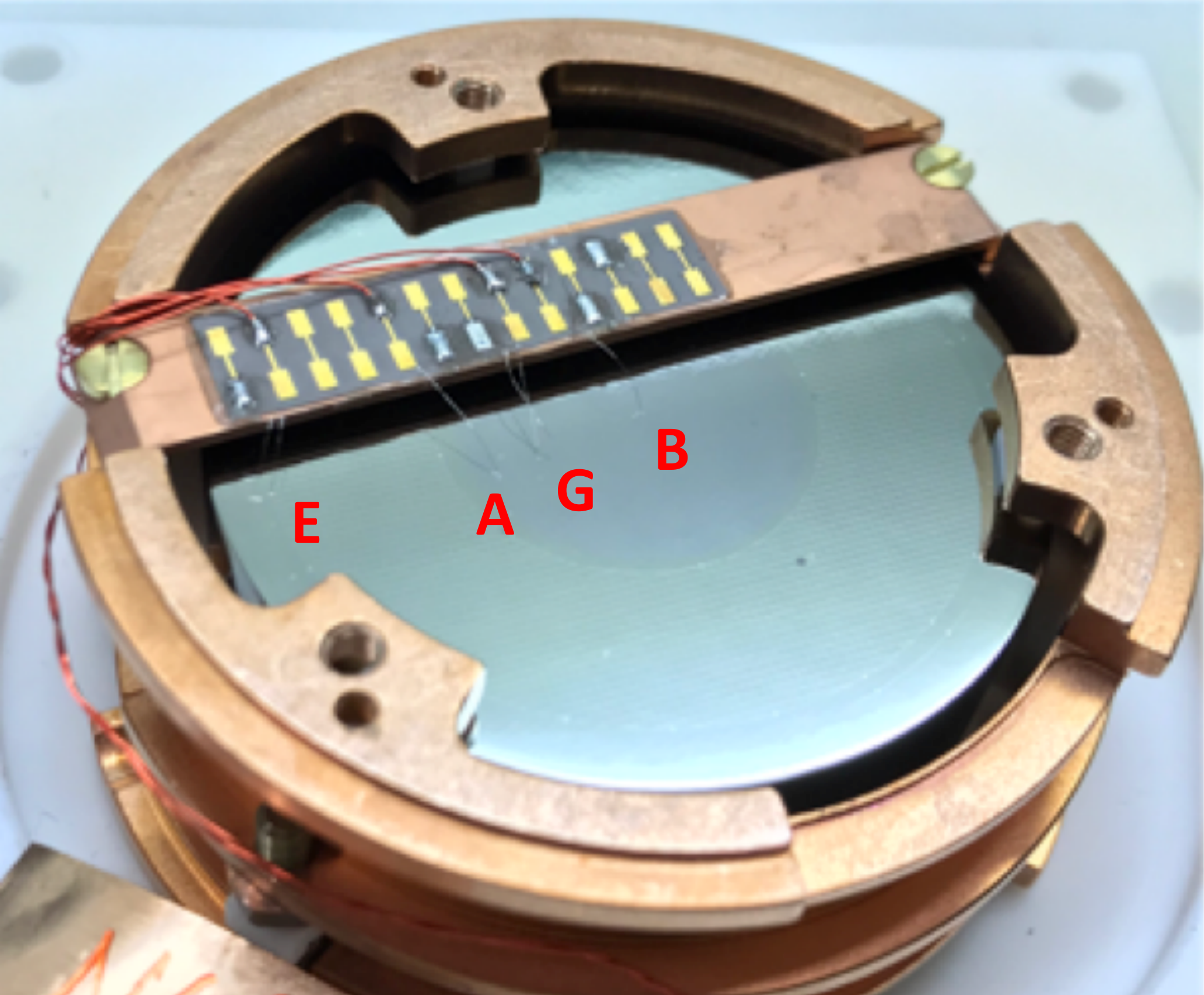}
\includegraphics[width=0.95\linewidth,trim= 0.0in 0.0in 0.0in 0.0in]{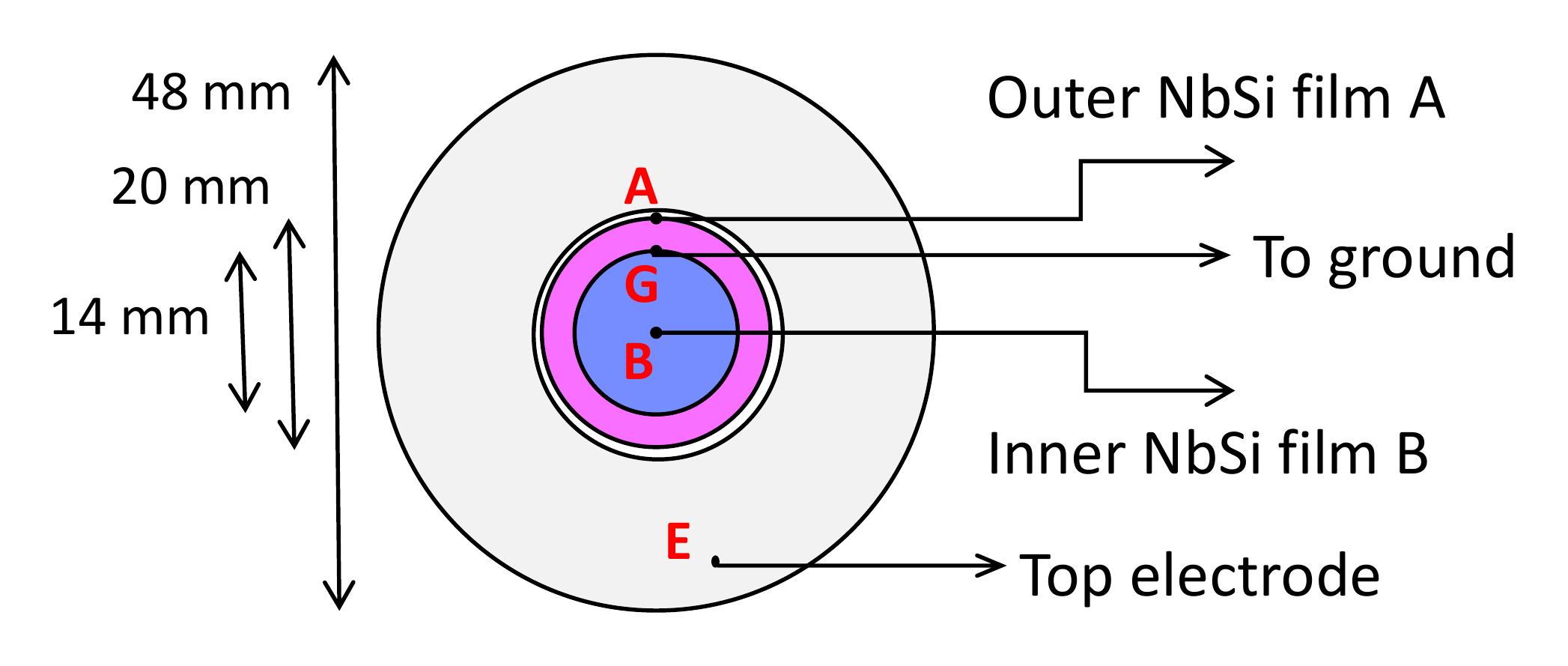}
\includegraphics[width=0.95\linewidth,trim= 0.0in 0.0in 0.5in 0.25in]{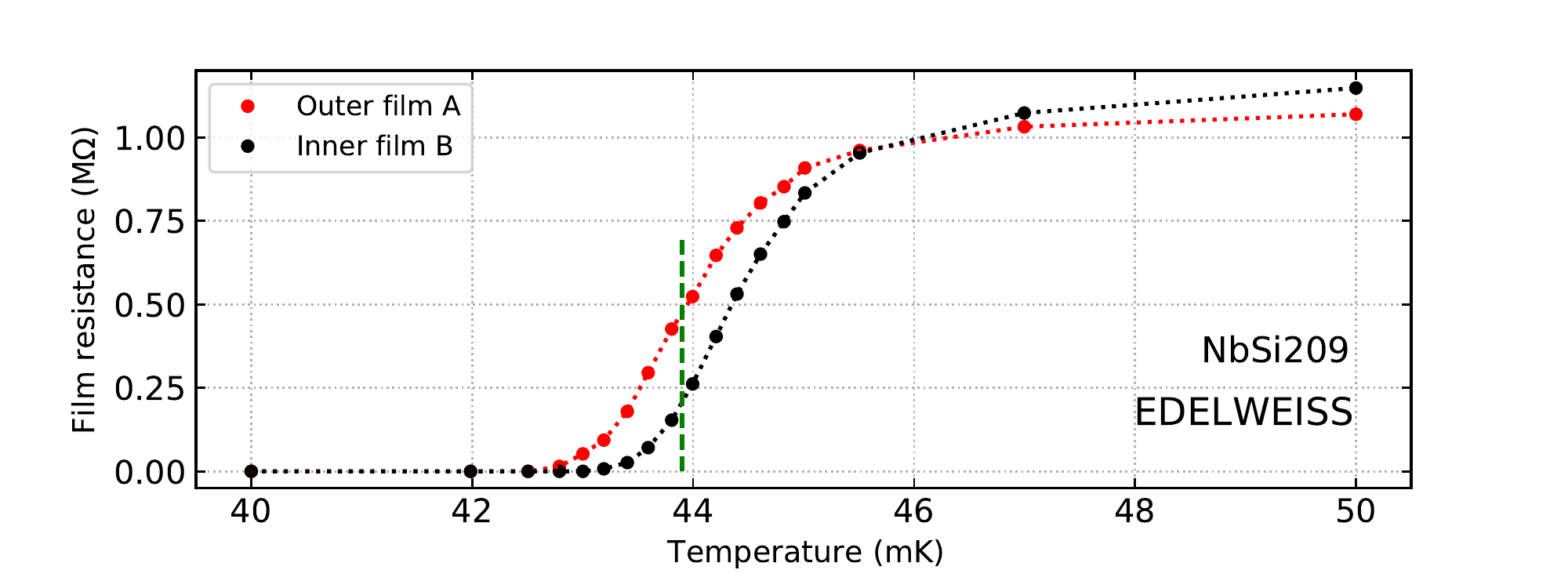}
\caption{Top: picture of top face of the NbSi209 detector, in its copper support. The diameter of the Ge cylinder is 48~mm. The darker grey circle at the centre of the top surface (20~mm diameter) is the NbSi film. 
Middle: geometry of the top face. As in the previous panel, the letters correspond to the electrical contact for the top electrode (E), the outer and inner extremities of the NbSi spiral resistor (A and B, respectively), and the point where it is connected to ground (G). 
Bottom: measured resistance of the two film halves as a function of temperature. The green vertical dashed line represents the operating temperature.}\label{fig-detector}
\end{center}
\end{figure}

\begin{table*}[t]
\centering
\setlength{\tabcolsep}{7pt}
\begin{tabular}{ |c|c|r|c|c|c|c|c|c|c| } 
 \hline
  Sampling & Electrode & Lifetime &  $^{71}$Ge & \multicolumn{6}{c|}{ Baseline resolution $\sigma$ } \\
  \cline{5-10} 
  (Hz) & bias (V) & (day) & activated &
   \multicolumn{3}{c|}{ in eV } & \multicolumn{3}{c|}{ in eV$_{ee}$ } \\
& & & & $\sigma_{outer}$ & $\sigma_{inner}$ & $\sigma_{phonon}$ & $\sigma_{ee}$ &$\sigma_{IonTot}$ & $\sigma_{IonBottom}$ \\
\hline
 &+66    &17.36 &Yes& 165 & 143 & 116 & 5.0 & 215 & 333 \\ 
 &$-$66    & 6.63 &Yes& 159 & 137 & 110 & 4.8 & 203 & 312 \\ 
 &+30    & 0.86 &Yes& 163 & 138 & 113 & 10.3 & 201 & 298 \\
 &$-$15 or +15& 5.57 &Yes& 151 & 131 & 102 & 16.9 & 204 & 304 \\
500 &+8     & 9.14 &Yes& 153 & 134 & 104 & 28.2 & 208 & 304 \\
 &0      & 0.51 &Yes& 166 & 143 & 112 & 112 & 199 & 297 \\
 \cline{2-10}
& $-$66 or +66& 50.63 & No & 138 & 139 & 103 & 4.5 & 209 & 325 \\
& $-$15 or +15 & 126.1 & No & 131 & 133 & 96 & 16.0 & 215 & 330 \\
\hline
\hline
 100000 & $-$66 or +66 & 1.73 & No &  &  & &  &  &  \\
      & $-$15 or +15 & 1.78 & No &  &  & &  &  &  \\
\hline
\end{tabular}
\caption{Summary of the data sets used in this paper. The data sets with the $^{71}$Ge activation were used for detector response measurements. The average baseline resolutions for the different energy measurements are listed for: the phonon energy measured by the outer and inner films ($E_{\mathrm{outer}}$ and $E_{\mathrm{inner}}$, respectively); their combination in units of keV ($E_{phonon}$) or in keV equivalent-electron $(E_{ee})$; the average of the two electrode signals ($E_{IonTot}$) and the bottom electrode signal alone ($E_{IonBottom}$). The systematic uncertainties on the resolution values are of the order of 2\%. The 100~kHz data were recorded in the presence of a $^{56}$Co gamma-ray source. The two longer data sets at $\pm$15~V and $\pm$66~V, recorded before and after the activation, were used for a DM search and a measurement of the Heat-Only population, respectively.}\label{tab-dataset}
\end{table*}

Fig.~\ref{fig-detector} shows a picture of the detector NbSi209 used in the present work. 
The absorber is a cylindrical germanium crystal with a height of 20~mm and a diameter of 48~mm, for a total mass of 0.20~kg. 
Aluminum grid electrodes are lithographed on the top and bottom flat surfaces.
In order to reduce the absorption of phonons, the top and bottom grids are made of 10~$\mu$m Al lines with a 500~$\mu$m pitch. 
This pattern ends at the outer 1~mm of the circumference which is covered by a plain Al surface. 
The bottom side of the detector is entirely covered by this electrode. 
A 20-mm wide circular area on the top side (middle panel of Fig.~\ref{fig-detector}) is not covered by Al but instead by a NbSi thin film~\cite{nbsi-ltd}. 
The film itself is a 100-nm thick spiral with a track width of 160~$\mu$m and a 40~$\mu$m gap between the tracks, for an aspect ratio of 10$^4$.
The spiral is separated in two halves of equal resistance via a wire bond contact at a radius of 7~mm connected to ground.
A heating resistor is glued on the bottom electrode in order to maintain the detector at a fixed temperature relative to the mixing chamber of the cryostat. 
The resistances of the two films as a function of the base temperature are shown on the bottom panel of Fig.~\ref{fig-detector}.
The critical temperatures for their transition from superconductor to a resistance of approximately 1~M$\Omega$ is 44 mK.
The critical temperatures of the two films differ by less than 0.5 mK.

The resistance of the films is continuously sampled using the standard EDELWEISS-III electronics for the readout of high-impedance signals~\cite{edwtech}.
The two films are read as two independent sensors.
They are each injected with a constant current switching sign at a rate of 500~Hz, and the corresponding voltage across each sensor is read via a biFET cooled at 100~K.

Due to the grounding of one side of both films, the electrode on the top side had to be kept at 0~V.
The bias applied to the bottom electrode could be varied from $-$66~V to +66~V.

The signals that are read out consist of the two phonon sensor channels, and the charge signal from the top and bottom electrodes.
The data of these four channels are digitized at a rate of 100~kHz.
In the standard acquisition mode, the data are processed online to extract the difference of the negative and positive current measurement of the films every 2~ms, resulting in a 500~Hz sampling rate written on disk.
For runs devoted to the measurement of the detailed phonon pulse shape, the data were stored on disk at the original 100~kHz rate.
This mode of operation is limited to a few special runs because it requires a factor 200 more space on disk space without any improvement in achieved energy resolution.

As in Ref.~\cite{nbsi-migdal}, the signal amplitudes were evaluated offline using measured pulse shapes and optimum filtering based on the noise spectra, evaluated hour-by-hour. 
Triggering and analysis efficiencies are determined with the same pulse simulation procedure.

The present analysis is based on data sets recorded during the cool-down that lasted from March 2019 to June 2020 described in Ref.~\cite{red30,nbsi-migdal}, during which the temperature of the detector support plate was regulated at either 20.7 or 22~mK.
The heater bias was set at a value that kept the film resistances at constant operating values.
The different data sets are listed with their lifetime in Table~\ref{tab-dataset}.

\section{Detector performance}

The ionization channel was calibrated using events recorded with the gamma-rays from a $^{133}$Ba source, as well as the K-shell events due to the activation of $^{71}$Ge inside the crystal obtained by exposing it to an AmBe neutron source~\cite{nbsi-migdal}.
The ionization gain is the same at 15 and 66~V, and remained constant over the entire cool-down. At 8~V, it is reduced by 4\% due to volume charge trapping effects.
The baseline resolution on the bottom electrode signal $\sigma_{IonBottom}$ varies between 297 and 333 eV$_{ee}$. The corresponding values for the signal derived from the average of the two electrodes $\sigma_{IonTot}$ varies between 199 and 215 eV$_{ee}$. 
 
The calibration of the signal of the phonon sensor in units of keV equivalent-electron (keV$_{ee}$) was also based on the $^{133}$Ba source and $^{71}$Ge K-shell activation measurement, as well as the phonon/ionization signal ratio of electron recoil events in the range from 5 to 200 keV.
The performance of the phonon sensors depends on the temperature of the crystal and on the currents applied to each phonon sensor.
These three parameters were systematically varied in order to find the conditions that optimize the baseline resolution on the combined phonon signal, obtained as the sum of the two sensor signals weighted by the square of the inverse of their resolution.
A broad optimum was found when the heater is set such that the inner and outer resistances are set to values of 200 and 450 k$\Omega$ (as shown by the green vertical dashed lines in the lower panel of Fig.~\ref{fig-detector}), while the currents are set to 3.5 and 2.0 nA, respectively.
The inner (resp. outer) film sensitivity at this operating point is 630 (370) nV/keV.
The resolutions obtained in the different data sets used in this work are listed in Table~\ref{tab-dataset}.
As in Ref.~\cite{nbsi-migdal}, periods where the hourly average of the resolution of the combined phonon signal $\sigma_{phonon}$ is above 140 eV are excluded from the present analysis.
With this cut, the average value of $\sigma_{phonon}$ for the data at 66~V used for the DM search is 103 eV.
The average values for the individual films are 139 and 138 eV for the inner and outer signal, respectively. Depending on noise conditions, these value can vary by as much as $\pm$25~eV.

The trigger on the inner phonon signal is identical to that of Ref.~\cite{nbsi-migdal}, with a plateau efficiency reached at energies above 1 keV (0.05 keV$_{ee}$ at 66~V).
The data quality cuts are also those of Ref.~\cite{nbsi-migdal}, with the exclusion of the cuts on the ionization signal not used here and on the phonon channel quality cut as we are interested to use events with non-standard pulse shapes). 
With the omission of these cuts, the plateau efficiency for events with full charge collection is 65.6\%. 

The dissipation of the work performed to drift the electrons and holes across the detector through a voltage bias difference $V$ adds a contribution $eNV$ to the thermal signal, where $N$ is the number of collected charge pairs (the so-called Neganov-Trofimov-Luke effect). 
This signal amplification also applies to athermal phonon signals~\cite{cdms-tes}.
The total phonon energy $E_{phonon}$ is converted in units of keV$_{ee}$ using the formula $E_{ee} = E_{phonon}/(1+e|V|/\epsilon_{\gamma})$, where $\epsilon_{\gamma}$~=~3~eV is the average energy per electron-hole pair in Ge and $e$ is the elementary charge.
The corresponding average resolution on $E_{ee}$ is $\sigma_{ee}$~=~4.5~eV$_{ee}$ at 66~V.
The resolutions at other bias voltage values are found in Table.~\ref{tab-dataset}.

\begin{figure}[htb]
\begin{center}
\includegraphics[width=0.99\linewidth,trim= 0.1in 0in 0.4in 0.3in]{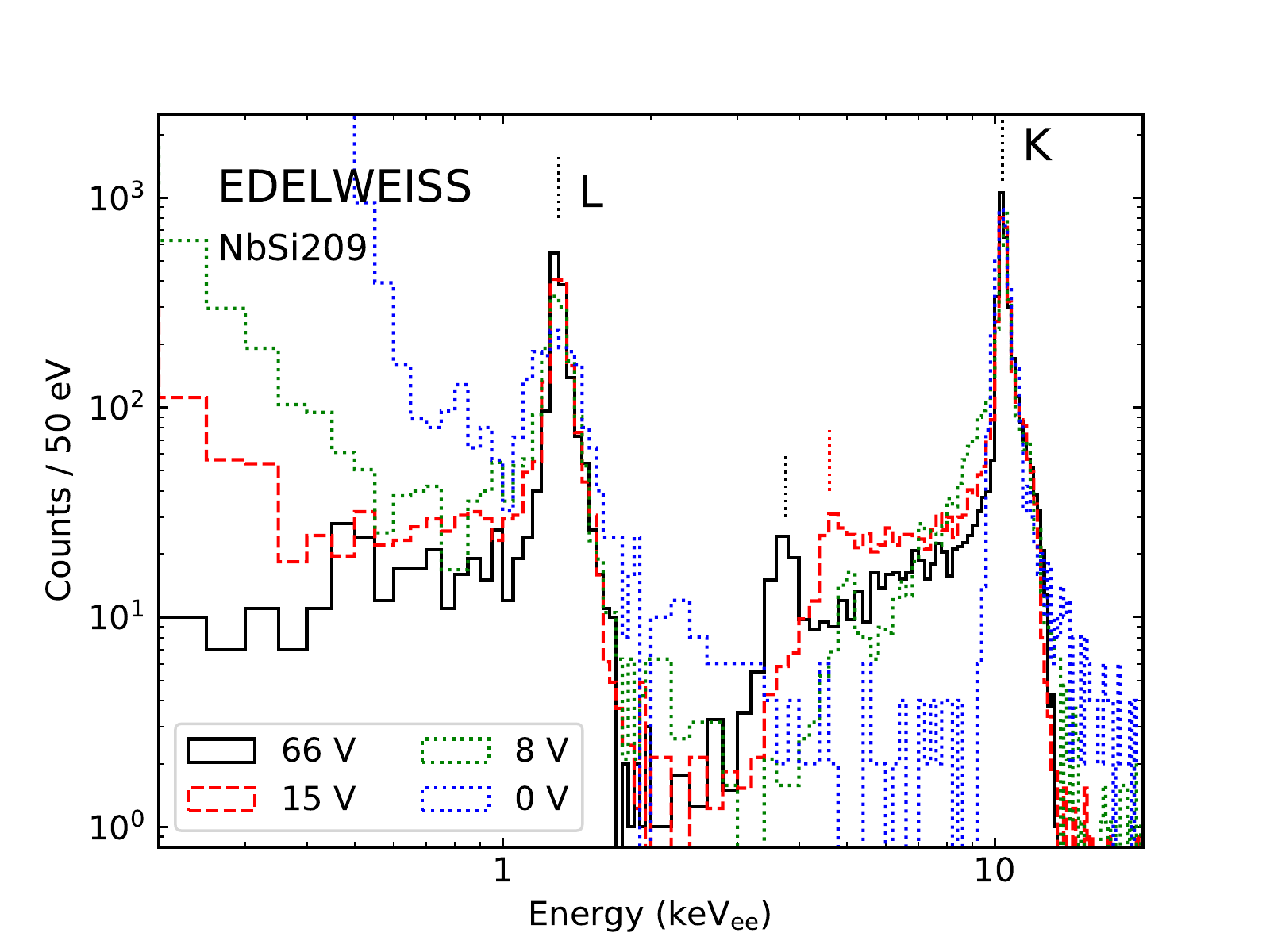}
\caption{Phonon energy spectra recorded at bias voltage of 0, 8, 15 and 66~V following the $^{71}$Ge activation of the crystal. The labels at 1.3 and 10.37 keV correspond to the position of the L- and K-shell peaks, respectively. The black and red vertical dotted lines at 3.75 and 4.61 keV represent the energy corresponding to 10.37 keV events where only 33\% of the charge is collected.}\label{fig-energyee}
\end{center}
\end{figure}

The most precise measurement of the detector performance has been done following the neutron activation of $^{71}$Ge in the detector volume~\cite{red30,nbsi-migdal}.
Fig.~\ref{fig-energyee} shows the phonon energy spectra, in keV$_{ee}$, recorded at different biases following this activation. 
The positions of the L- and K-shell peaks at 1.30 and 10.37 keV (as well as the M-shell peak at 0.16 keV observed at 66~V in Ref.~\cite{nbsi-migdal}), made possible a precise determination of the non-linearity of the phonon channel.
It is less than 5\% over the phonon energy range from 1 to 500 keV. 
The observed phonon gains at bias voltages of 8, 15, 30 and 66 V are all compatible with those expected from the $(1+e|V|/\epsilon_{\gamma})$ dependence of the NTL amplification.

The number of events in the tails on the left-side of the K- and L-shell peaks allows to quantify the fraction of volume affected by incomplete charge collection. 
The peaks at 3.75 and 4.61 keV$_{ee}$ labelled on Fig.~\ref{fig-energyee} correspond to a loss of 67\% of the charge at 66 and 15~V, respectively.
This is a likely value for events occurring at the bottom of the outer cylindrical surface of the detector according to simulation of the Ramo potential.
These important losses are due to the absence of an electrode on the outer cylindrical surface and the proximity of the grounded copper casing.
The tails extending to the right of the peaks are barely visible on that figure, but their effect will be made more prominent in the study of the excesses of athermal phonon signals in either one of the two films presented in the following sections.

Nevertheless, at 66~V most K-shell events (63\%) are centered at 10.37 keV$_{ee}$ within a gaussian peak with a $\sigma$ value of 180 eV$_{ee}$.
These were the events selected for the DM search of Ref.~\cite{nbsi-migdal}.
The following study will look in more details into the contrasting behavior of this population and all others forming the low and high energy tails of the phonon energy distribution.

\section{Data selection}

To better understand the phonon spectra associated to different populations, we will use the ionization signals to separate the events according to their different position inside the detector.

\begin{figure}[htb]
\begin{center}
\includegraphics[width=0.99\linewidth,trim= 0.1in 0in 0.4in 0.3in]{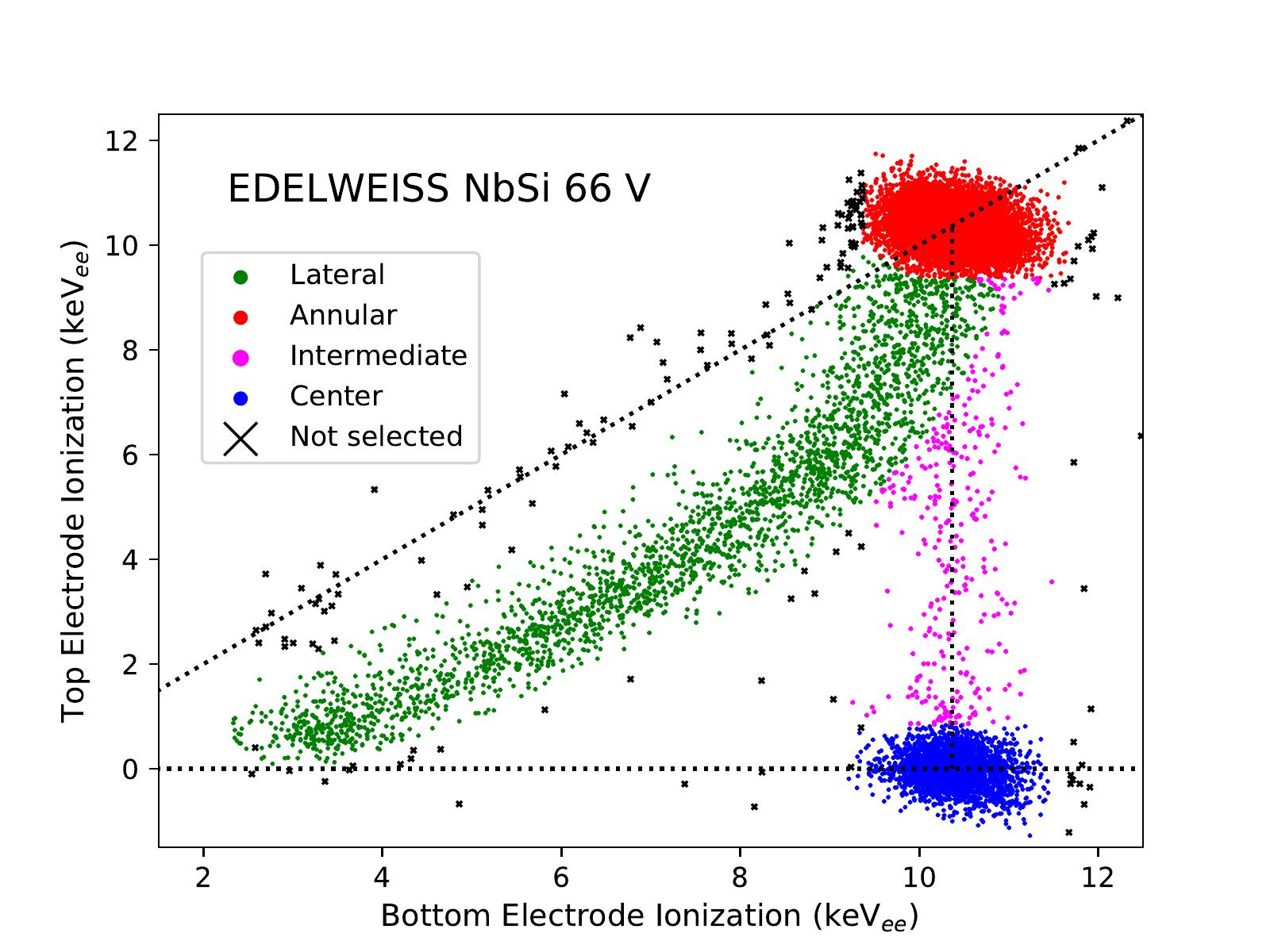}
\includegraphics[width=0.95\linewidth,trim= 0.0in 0.0in 0.0in 0.0in]{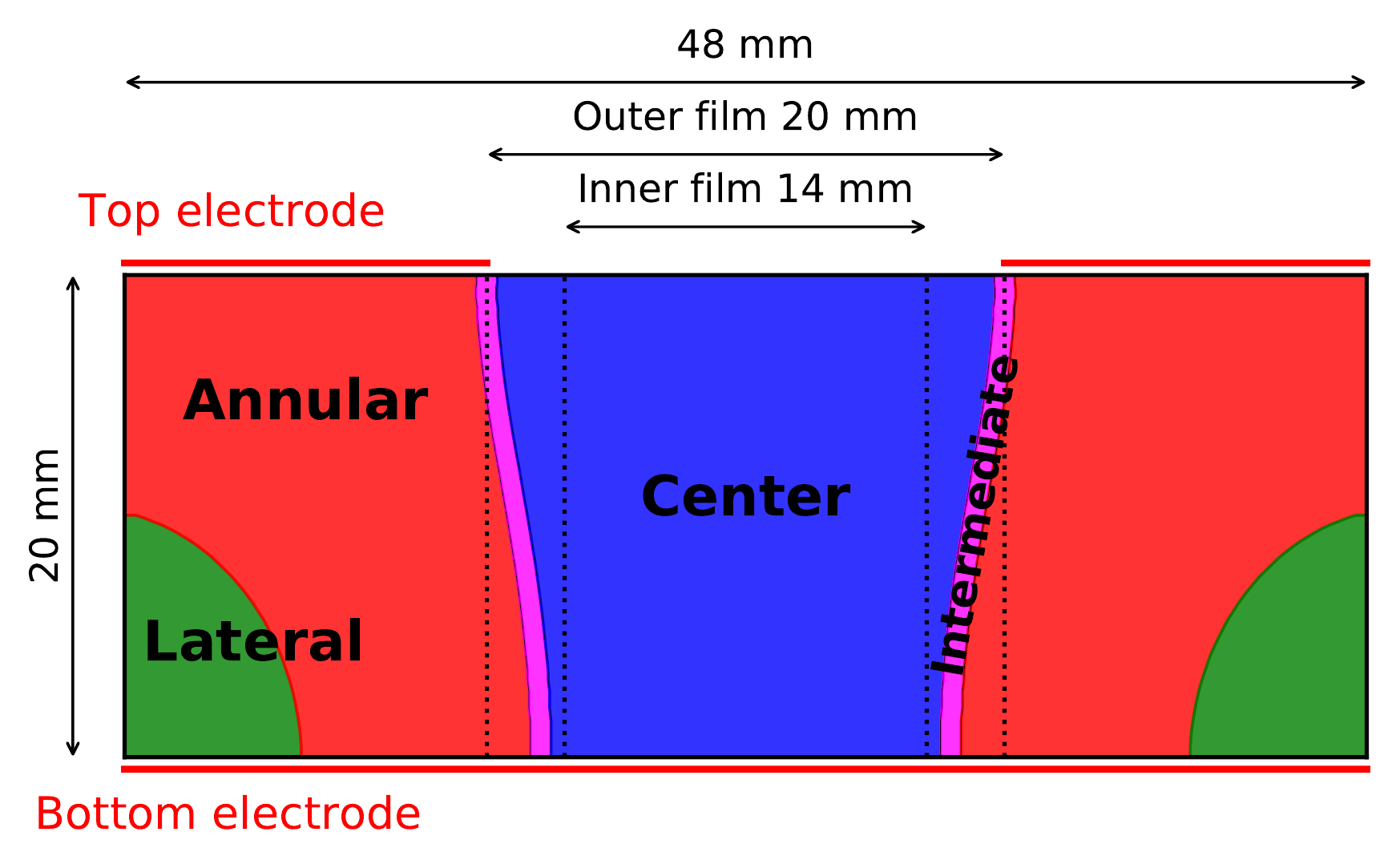}
\caption{Top: ionization signals on the top and bottom electrodes recorded at 66~V after $^{71}$Ge activation. The color code corresponds to the selection on the ionization channels described in the text.
Bottom: scheme of the geometry of the NbSi films and the electrodes. The colors represent the different volumes bounded by the field lines described in the text.  }\label{fig-ionsel}
\end{center}
\end{figure}

The top panel of Fig.~\ref{fig-ionsel} shows the ionization signals on the top electrode ($E_{top}$) versus the bottom one ($E_{bottom}$) for events recorded at $\pm$66~V after the $^{71}$Ge activation, with bottom signals between 2.3 and 12 keV$_{ee}$.
Most events on that figure can be attributed to K-shell conversions, with the exception of the small number of events along the $E_{top}=E_{bottom}$ diagonal below 9 keV$_{ee}$, marked as black crosses.
The different colors of the K-shell events shown in Fig.~\ref{fig-ionsel} correspond to four selection cuts based on their $(E_{top},E_{bottom})$ values.
These selections are labeled \emph{lateral, annular, intermediate} and \emph{center}, for reasons that will be explained later.
The fractions of events in each category are found in Table~\ref{tab-fraction}.
The blue \emph{center} population is characterized by a complete collection of a 10.37 keV signal on the bottom electrode, and none on the top one. 
The red \emph{annular} events have a 10.37 keV peak on both top and bottom electrodes.
An \emph{intermediate} population, in magenta, stretches between the previous two populations.
The fourth population, in green, consists in a tail extending from the annular region down to $(E_{bottom},E_{top})$ $\sim$ (3.5, 0.5) keV. This \emph{lateral} population overlaps partially the magenta \emph{intermediate} one. 
Considering the relative strength of the four population in Table~\ref{tab-fraction},
the contamination of the \emph{lateral} population with \emph{intermediate } events should be small.

\begin{table}[htb]
\setlength{\tabcolsep}{3pt}
\begin{center}
\begin{tabular}{ |r|c|c|c|c| } 
 \hline
 Data  & \multicolumn{4}{c|}{Event category } \\ 
 \cline{2-5}
  set& Lateral & Annular & Intermediate & Center \\ 
\hline
 +66~V & 19.6 $\pm$ 0.5 & 60.9 $\pm$ 0.6 &  2.7 $\pm$ 0.2& 16.9 $\pm$ 0.5\\ 
 -66~V & 17.8 $\pm$ 0.3 & 63.7 $\pm$ 0.4 &  2.1 $\pm$ 0.1& 16.4 $\pm$ 0.3 \\ 
  +30~V & 18.6 $\pm$ 0.7 & 62.4 $\pm$ 0.9 &  3.9 $\pm$ 0.3& 15.1 $\pm$ 0.6\\
  $\pm$15~V & 22.1 $\pm$ 0.4 & 57.8 $\pm$ 0.3 &  4.2 $\pm$ 0.2& 15.8 $\pm$ 0.3\\
   +8~V & 31.8 $\pm$ 0.6 & 46.0 $\pm$ 0.6 & 13.4 $\pm$ 0.4&  8.8 $\pm$ 0.3\\
 \hline
\end{tabular}
\caption{Fraction of K-shell events in the different event categories in Fig.~\ref{fig-ionsel}. The errors are statistical only. The bias sign corresponds to $V_{top}-V_{bottom}$, i.e. electrons are collected on the top electrode when $V>0$.}\label{tab-fraction}
\end{center}
\end{table}

The origin of these four populations, and the reason why $E_{top}$ $<$ $E_{bottom}$ for some of them, can be understood by inspecting the diagram in the bottom panel in Fig.~\ref{fig-ionsel}, where the color regions are defined by the geometry of the lines calculated with a COMSOL\textsuperscript{\textregistered}~\cite{comsol} simulation of the detector.
All field lines connect to the bottom electrode, as it is the only one biased relative to the grounded copper cover.
There are two cases where the other end of the field lines does not reach the top electrode: some field lines intersect the surface covered by the NbSi sensor, while others exit the detector via its cylindrical lateral surface, which is not covered by an Al electrode. 
The blue \emph{center} region is defined by all field lines connecting the bottom electrode to the 20~mm diameter region covered by the two NbSi films.
In this case, all the charge drifting upward will be collected in the NbSi film, also set at ground.
These charges are not read out, and therefore $E_{top}$ is zero, resulting in the blue population on the top panel of Fig.~\ref{fig-ionsel}.
In the other case where electrons or holes end their drift on the lateral surface, the signals induced in both electrodes must be calculated using the Shockley-Ramo theorem~\cite{shockley,ramo,ramo-polar,ramo-ge}.
This will induce a reduced, but non-zero, signal on the top electrode.
The bottom electrode signal will also be degraded, while remaining greater than the top one due to the field asymmetry. 
For example, the green \emph{lateral} area on the bottom panel of Fig.~\ref{fig-ionsel} corresponds to the region where the COMSOL\textsuperscript{\textregistered} calculation predicts that the signal induced in the bottom electrode corresponds to less than 95\% of the total charge. 
This mechanism provides an explanation for the green \emph{lateral} population on the top panel of Fig.~\ref{fig-ionsel}, where the bottom electrode signal deviates from the nominal 10.37 keV$_{ee}$.
The process of NTL amplification is also affected, and this population is also at the origin of the tails on the left-hand side of the L- and K-shell peaks observed in Fig.~\ref{fig-energyee}.

This simple model does not take into account the effect of anisotropy, charge repulsion and bulk trapping on the electron and hole drift~\cite{anisotropy}, that can be very important in particular at low field.
Such effects can explain the origin of the small magenta population on the top panel of Fig.~\ref{fig-ionsel}, where the reduction of the top signal is not accompanied by a degradation of the bottom signal. 
In this case, the missing top signal has disappeared in the NbSi film, with the remaining share of charge on the electrode varying almost almost uniformly between 0 and 100\%.
Such charge-sharing effects between adjacent electrodes have been observed before in EDELWEISS detectors~\cite{charge-sharing}, and empirically described in terms of a simple model where the number of affected events depends on a characteristic radius that averages the effects of all mechanisms that give rise to a lateral dispersion of the charges relative to the field lines, averaged over all event depths.
In Ref.~\cite{charge-sharing}, typical values of the lateral spread at a field of 2 V/cm$^{2}$ are $\sim$2~mm.
This would result in a fraction of \emph{intermediate} events of the same order than what is measured at 8~V (13.4\%, see Table~\ref{tab-fraction}).
At higher biases the measured fraction decreases, as would be expected from the reduction of anisotropy and trapping effects as the electric field increases.
In this simple model, an average lateral spread of 0.4~mm would yield an \emph{intermediate} population fraction of 2.5\%, close to the observed values at $-$66~V and $+$66~V, and corresponding to the magenta volume depicted in the bottom panel of Fig.~\ref{fig-ionsel}.

Accurate population fraction predictions would require a detailed modelling of the charge drift that is beyond the objective of this paper.
Here, it is sufficient to note that the \emph{center, intermediate, annular}, and \emph{lateral} event selections, based on the ionization signals, correspond to interactions occurring at increasing values of radius inside the detector volume. 
These selections thus provide event samples that will enable the study of the dependence of the phonon signals on the lateral position of the interaction within the detector volume.

The full identification of the four event populations can only be performed using K- or L-shell events.
Their differences in phonon amplitudes are studied in Sect.~\ref{sect-500Hz} using the 40.1-day data sample recorded following the $^{71}$Ge activation of the detector listed in Table~\ref{tab-dataset}.
But first, in order to properly interpret the signal amplitudes in terms of phonon energy, a good understanding of the phonon pulse shape is required.
For this, the results of a detailed pulse shape study will be presented in Sect.~\ref{sect-100kHz}. It uses a dedicated 3.5-day data set (see Table~\ref{tab-dataset}) recorded on disk at 100~kHz (instead of the coarser 500~Hz sampling of the larger 40.1~day data sample). 
In addition, the precision on the measurements of the individual event pulse shapes is improved by using  higher energy events (10 to 200 keV$_{ee}$ Compton scatters) from a $^{56}$Co gamma-ray source.
For this first study, in the absence of a monoenergetic K-shell line, the essential feature of the position-dependence of the pulse shape will be demonstrated by separating the \emph{center} events from the three other categories by using a cut on $E_{IonTop}<2\sigma_{IonTop}$. 

\section{Position dependence of the phonon pulse shape}\label{sect-100kHz}

The data analysis of Ref.~\cite{nbsi-migdal} assumed a standard pulse shape for all events. 
This assumption was correct given the event selection used in that work, which rejected events with significant difference between the $E_{inner}$ and $E_{outer}$ phonon signal amplitude. 
These events tended to be poorly described by the standard pulse shape.
The investigation of this effect motivated the collection of a relatively large sample data recorded at 100~kHz.

\begin{figure}[htb]
\begin{center}
\includegraphics[width=0.99\linewidth,trim= 0.0in 0.2in 0.4in 0in]{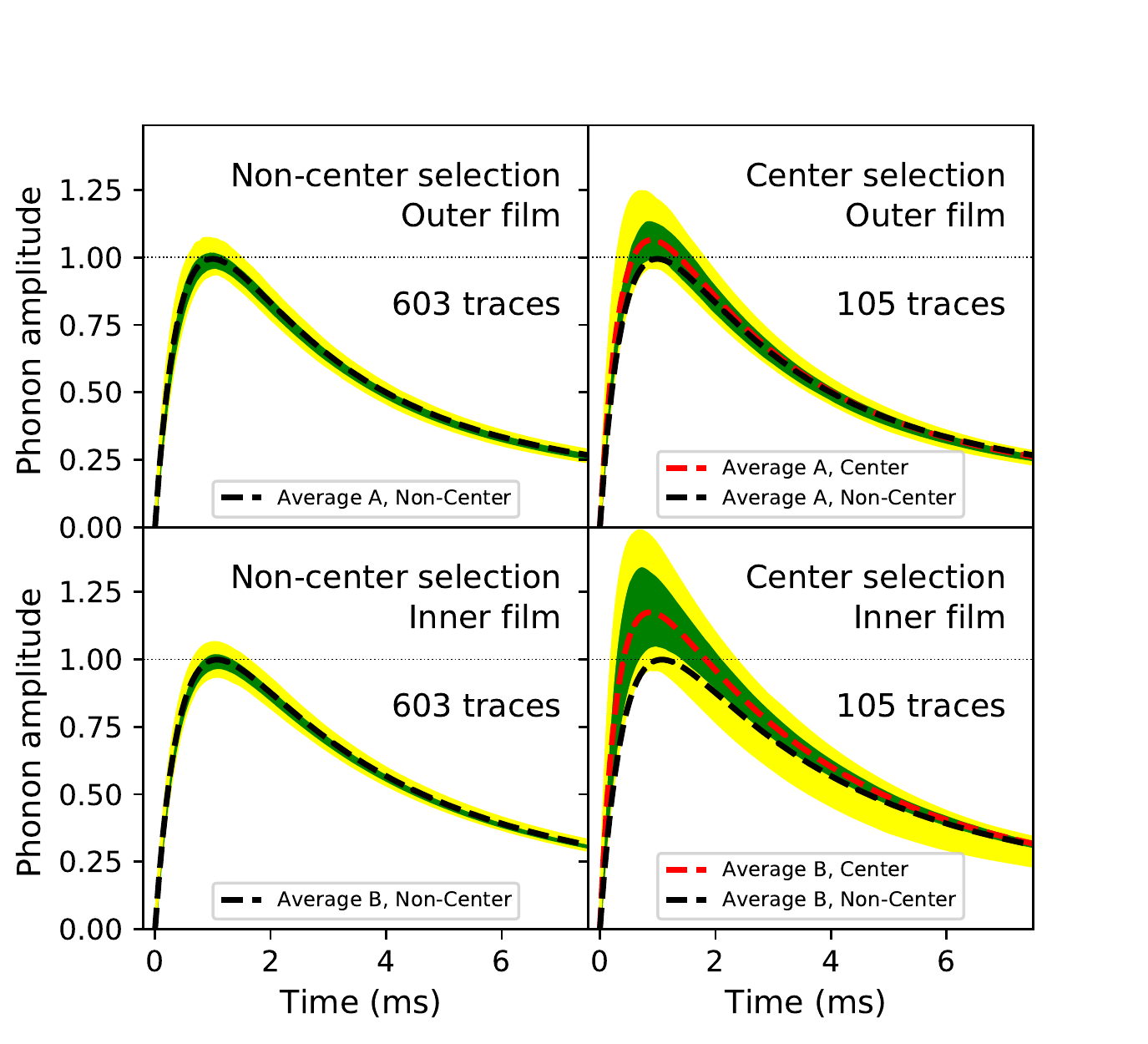}
\caption{Dispersion of the observed phonon pulse shape, normalized to the bottom electrode ionization signal and accounting for the NTL amplification, for events collected at both $-$66~V and +66~V. 
Top: outer film signal. Bottom: inner film signal. Right: \emph{center} selection ($E_{IonTop}<2\sigma_{IonTop}$). Left: \emph{non-center} selection, \emph{i.e} events rejected by the \emph{center} selection.
The black and red dashed lines are the average pulses for the \emph{center} and \emph{non-center} selections, respectively. The green and yellow bands cover 68\% and 95\% of the observed traces, respectively.}
\label{fig-100kHz}
\end{center}
\end{figure}

The phonon pulse shape of events in the 100~kHz sample corresponding to ionization events with energies $E_{IonBottom}$ between 20 and 200 keV$_{ee}$ were fitted by a pulse model with one exponential rise time constant $\tau_0$ and three independent time constants $\tau_1$, $\tau_2$ and $\tau_3$.
The events with $E_{IonBottom}$ greater than 53 keV$_{ee}$ were removed from the $\pm$66~V sample, to account for the onset of saturation effects observed for phonon energies above 1.2 MeV.
For each event, the phonon signal is divided by the measured ionization amplitude.

The results of the fit of 708 pulses recorded at $\pm$66~V are shown in Fig.~\ref{fig-100kHz}. 
The dispersion of the pulse shapes for events rejected by the \emph{center} selection $E_{IonTop}<2\sigma_{IonTop}$ is very small.
There is thus very little pulse shape variation for all \emph{non-center} events, corresponding to the three \emph{intermediate}, \emph{annular} and \emph{lateral} categories.
The RMS dispersion near the pulse maximum is less than 3\%.
Given this reproducibility, for convenience, all phonon amplitudes for a given film are expressed here in units of the maximum of the height of the average \emph{non-center} pulse shown in Fig.~\ref{fig-100kHz}. 
Considering that this selection corresponds to more than 80\% of the detector volume, this small dispersion implies a very small dependence and consequently proves that the signal is dominated by ballistic and/or thermal phonons. 
The contribution of short-range ($<$1~cm), high-energy ($>$ THz) phonons is negligible.
With this selection, the fit parameters for both films are consistent with a risetime $\tau_0$~=~0.43~ms, a strong (75\% amplitude) fast decay with $\tau_1$~=~3.0~ms and 2.2~ms for the inner and outer films, respectively, and two smaller slower components with $\tau_2$~=~11~ms (20\%) and $\tau_3$~=~111~ms (5\%).

In contrast, the average shape of \emph{center} events differs systematically from that of \emph{non-center}, with substantially larger event-by-event fluctuations. The risetimes are faster and the amplitudes are systematically larger than those of the average \emph{non-center} pulse.
The fluctuations are larger on the outer film than on the inner film.
The deviation from the average \emph{non-center} pulse is maximal in the first 1~ms and disappears after 5~ms.

\begin{figure}[htb]
\begin{center}
\includegraphics[width=0.99\linewidth,trim= 0.2in 0in 0.5in 0in]{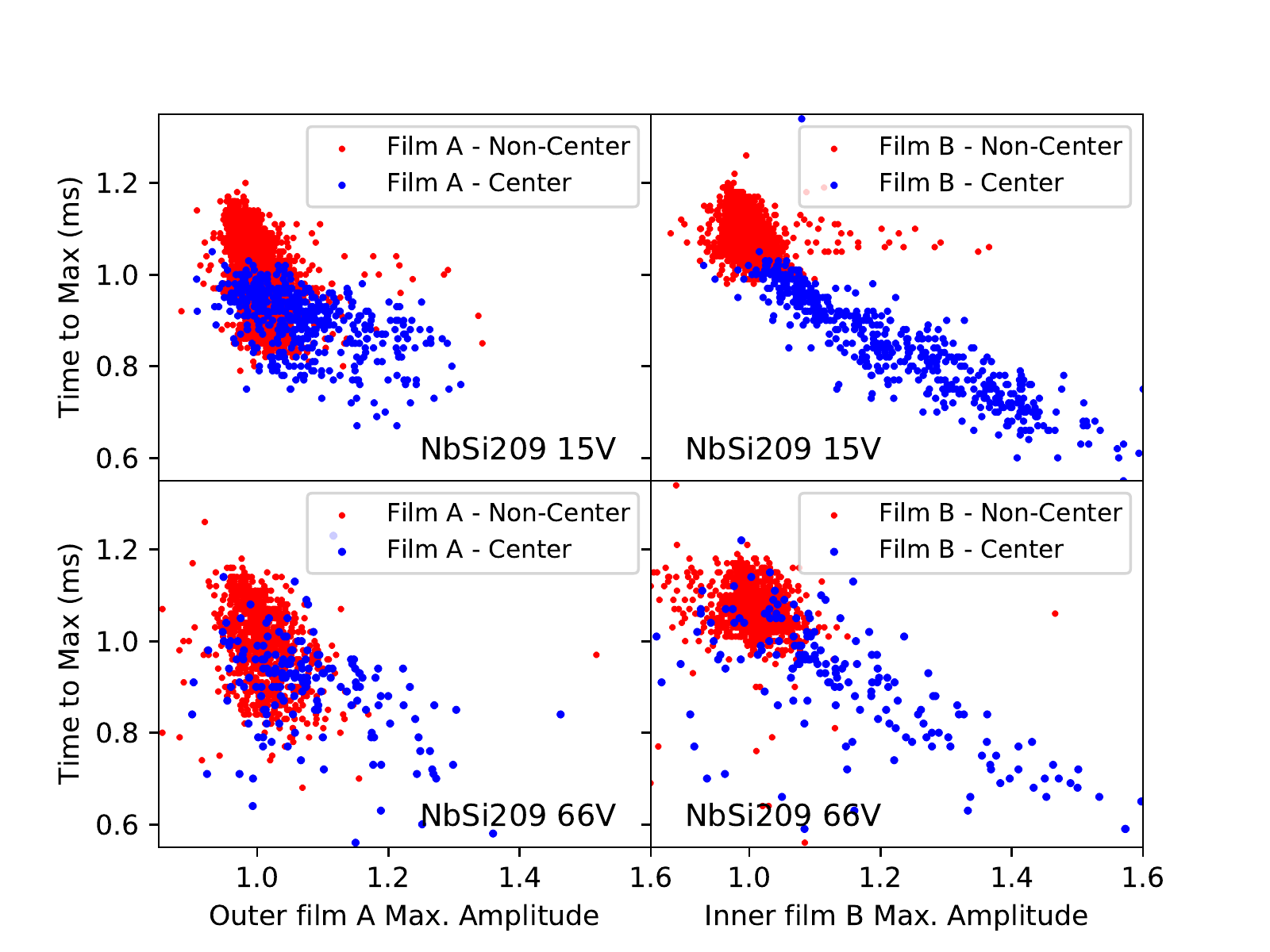}
\caption{Time to reach the maximum of the pulse versus the maximal amplitudes of NbSi film pulses recorded at 15 and 66~V. The pulse amplitudes are normalized to the signal on the bottom electrode.}\label{fig-timeconst}
\end{center}
\end{figure}

The same effect is observed in the $\pm$15~V data.
Fig.~\ref{fig-timeconst} compares the correlation between the time to reach the maximum ($t_{max}$) and the maximum of pulses ($H_{max}$) at $\pm$15 and $\pm$66~V for \emph{center} and \emph{non-center} events.
The \emph{non-center} populations at both biases show little dispersion in $t_{max}$ and $H_{max}$.
For \emph{center} events at both biases, the excess in amplitude reaches up to 40\%, with a correlated reduction of $t_{max}$.

In Fig.~\ref{fig-timeconst}, the largest values of $H_{max}$ ($\sim$1.4) on the inner film for \emph{center} events are associated with  $H_{max}$ $\sim$ 1 on the outer film. 
And vice-versa, the events with the largest values of $H_{max}$ ($\sim$1.2) on the outer film are in turn associated with $H_{max}$ close to 1 on the inner film. 
The excess amplitude observed for \emph{center} events mostly occur on one film at the time. 
This, together with the fact that no excesses are observed when the events occur away from the \emph{center} region, strongly suggests that the excess is due to the detection of nearby short-range, high-energy phonons before their down-conversion to lower-energy ballistic phonons.

The $\sim$0.5~ms risetime of the additional phonon component cannot be ascribed to the short lifetime of these primary phonons, as these are orders of magnitude faster \cite{cdms-phonontime}.
Propagation and relaxation effects inside the NbSi film itself may play an important role in determining the observed signal risetime.

A potential cause for the presence of an extra signal component in one film at the time for \emph{center} events could be that the arrival of the drifted charges inside a NbSi film alters its properties.
For example, the energy released by the recombination of the charges as they reach the film could lead to an additional signal.
However the amplitude of this components would be proportional to the number of charges arriving in the film, which is the same at 15~V and 66~V.
This is not consistent with the observation of similar excesses of 40\% at both bias values, since an excess with a fixed absolute energy would represent a smaller fraction of the normal phonon signal as it is amplified by a factor 3.83 between 15 and 66~V because of the NTL effect.
The fact that the inner film amplitude fluctuations extend up to a factor 1.4 at both 15 and 66~V indicates that the size of the excess scales with the strength of the total phonon signal (or at least its dominating NTL component), and not with the number of collected charges.

\section{Position, voltage and energy dependence of excess phonon signal}\label{sect-500Hz}

The data sets recorded at 500~Hz after the $^{71}$Ge activation (see Table~\ref{tab-dataset}) were analyzed in order to study excess phonon signals this time with a better diagnostic on the event location (using the K-shell peak events) and with higher statistic samples recorded at different biases.
The slower sampling frequency severely reduces the sensitivity to the details of the pulse shape. However, it was found that the amplitude ratio of the inner and outer films $r$ = $E_{inner}/E_{outer}$ provides a very good estimator of the strength of the extra component described in the previous section.

\begin{figure}[tbh]
\begin{center}
\includegraphics[width=0.99\linewidth,trim= 0.2in 0in 0.6in 0in]{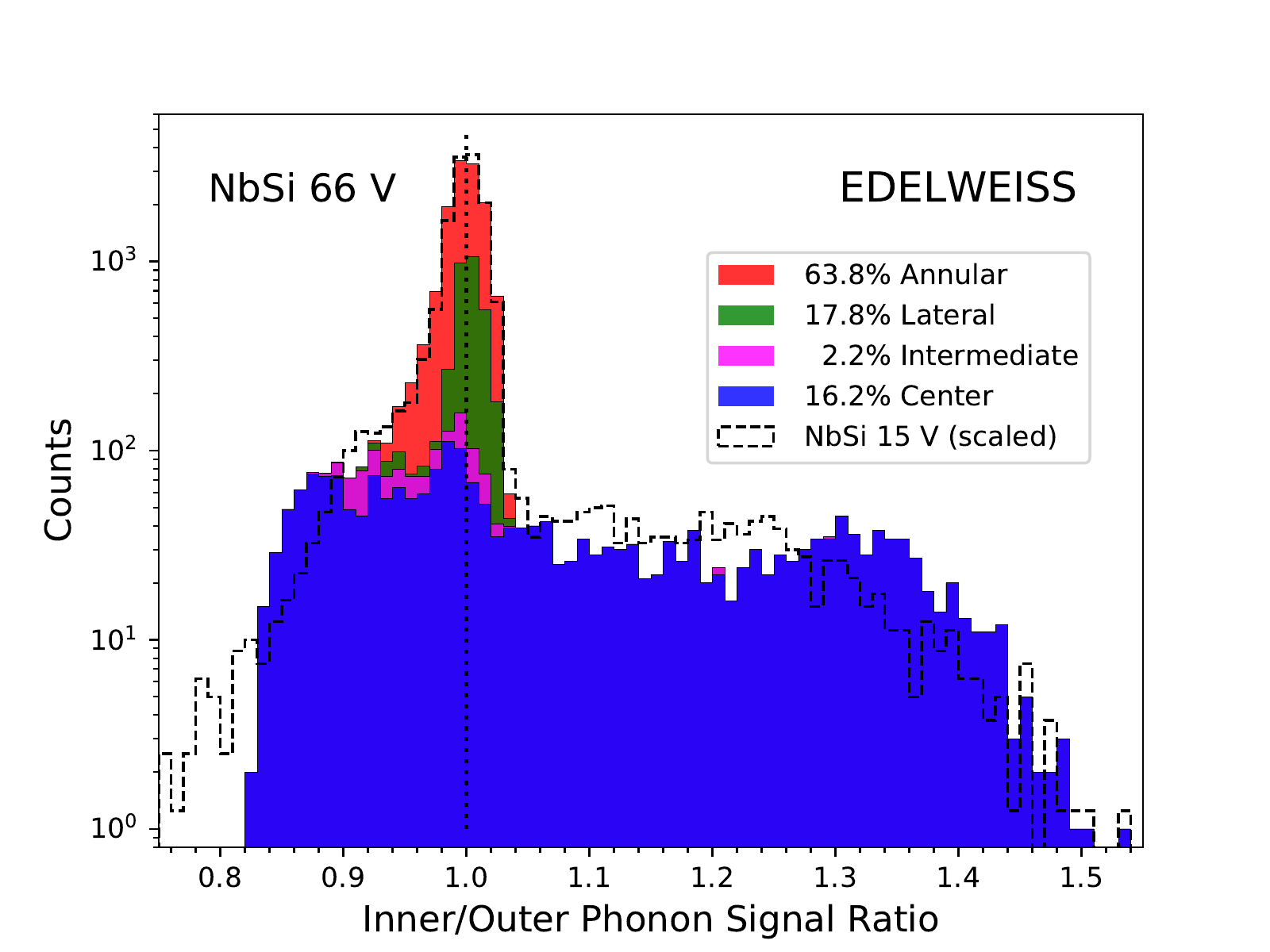}
\caption{Ratio $r$ of the amplitude of the inner and outer film phonon signal for the different event categories of Fig.~\ref{fig-ionsel}. 
The categories are shown using the same color code.
For comparison, the dashed line represents the distribution recorded at 15~V.} \label{fig-ratios}
\end{center}
\end{figure}

Fig.~\ref{fig-ratios} shows the $r$ distributions of the four selections of K-shell events defined in Fig.~\ref{fig-ionsel} for the data sample recorded at $-$66 and $+$66~V.
The histogram color code for the four event categories follows that of Fig.~\ref{fig-ionsel}.
The $r$ values for events in the \emph{annular} and \emph{lateral} selections are centered at one, with a small dispersion.
In contrast, a majority of events with the \emph{center} selection have $r$ values between 1.0 and 1.4, \emph{i.e.} close to the maximum values of inner film amplitude observed in Fig.~\ref{fig-timeconst} with a similar selection.
The amplitudes derived from the 500~Hz data are thus quite sensitive to the presence of the fast extra component described in the previous section.
The loss of information on the pulse shape is compensated with the improved energy resolution brought by the optimal filter technique used to evaluate a global pulse amplitude, compared with the uncertainties resulting from the multi-parameter pulse shape fit performed in the previous section. 
This advantage is particularly important in this section devoted to the study of low-energy K- and L- shell events.

\begin{figure}[tbh]
\begin{center}
\includegraphics[width=0.99\linewidth,trim= 0.2in 0in 0.5in 0in]{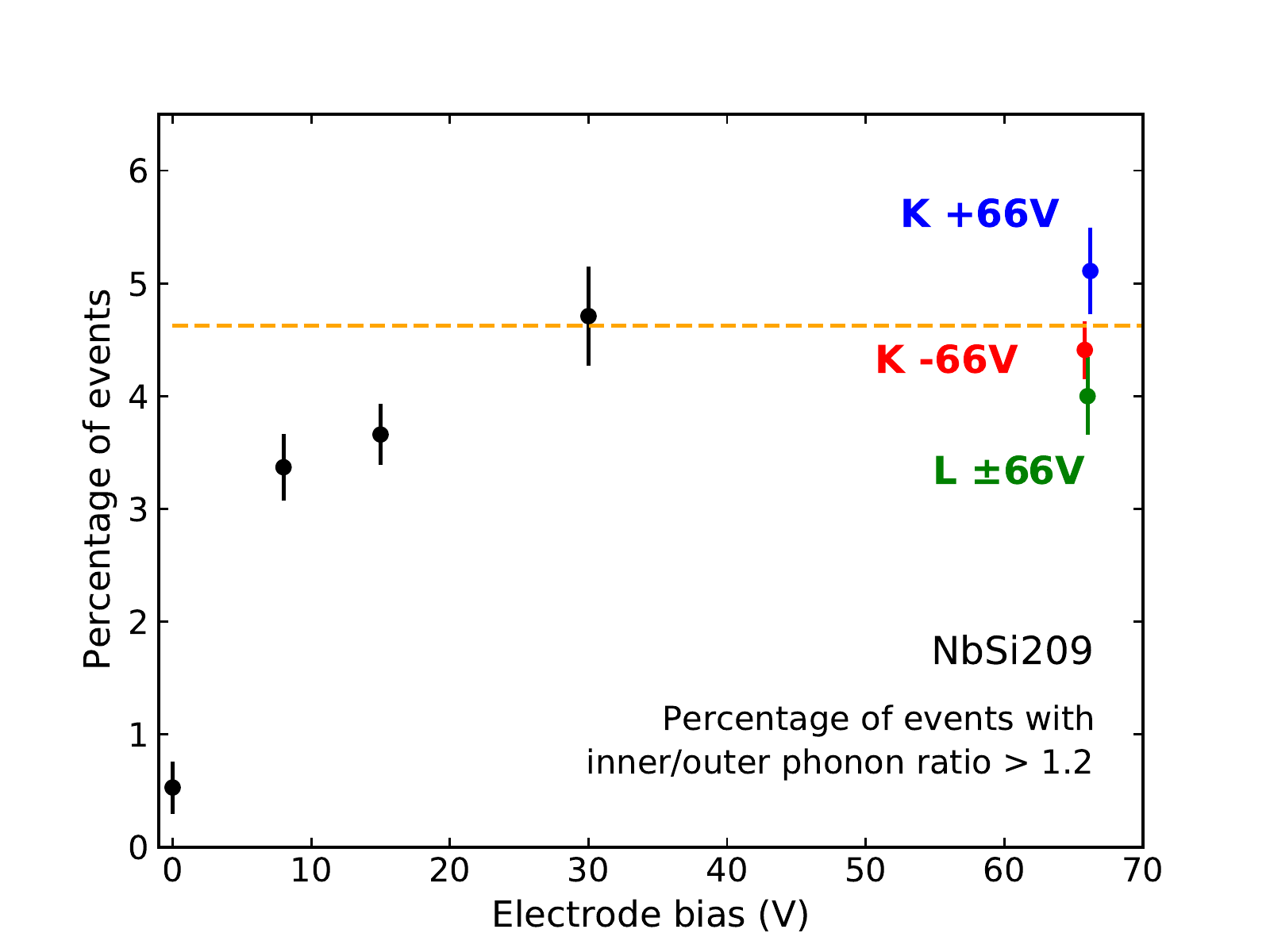}
\caption{Fraction of events with an inner/outer phonon ratio $r$ above 1.2 for K-shell at different electrode bias, and for L-shell events.}\label{fig-fraction}
\end{center}
\end{figure}

The large $r$ value fluctuations for the \emph{center} selection are almost entirely due to an excess of $E_{inner}$ signal relative to the nominal 10.37 keV$_{ee}$ value, while the $E_{outer}$ signals are much less affected.
For instance, the distribution in $E_{outer}$ of the events with $r>1.2$ in Fig.~\ref{fig-fraction} is observed to be a Gaussian centered at 95\% of the nominal position and with a $\sigma$ of 4.4\%, meaning that the anomalous $r$ values are essentially due to $E_{inner}$ excesses ranging from 15\% to 35\%.
The concentration of the excess in only one of the two films strongly suggests that phonons contributing to it have mean free paths that are small compared to the $\sim$mm size of the films.
The 4.4\% dispersion on the $E_{outer}$ signal opens the possibility to use this variable as an energy estimator for the normal phonon component of events with large $r$ values, once corrected for the 95\% bias in amplitude. 

A similar behavior, but this time exchanging the role of the inner and outer film, is also observed for the small (2\%) population of \emph{intermediate} events (in magenta in Fig.~\ref{fig-ratios}).
These events tend to have $r$ values below 1, and in this case it is the $E_{inner}$ value that remains close to the nominal one while the $E_{outer}$ values are in excess.

A short mean free path must also be invoked to explain that in Fig.~\ref{fig-ratios}, there are no \emph{intermediate}, \emph{annular} or \emph{lateral} events (representing (83.2 $\pm$ 0.5) \% of the total population) with $r$ $>$ 1.05, while it is the case for (45 $\pm$ 1)\% of the \emph{center} events. 
This means that the phonons at the origin of large excesses regularly occurring on the inner film in the case of \emph{center} events are never observed when their point of origin is anywhere inside the \emph{annular} region that is only distant by a mere 3~mm (\emph{i.e.} the width of the outer film).

The excess signal in a given film thus seems to be associated \emph{i)} with the proximity of the interaction to a region covered by that film and \emph{ii)} with phonons having mean free paths at the mm scale or below.

The existence of strong proximity effects for a large fraction of \emph{center } events comes as a violent contrast to the rather uniform response of the two films for events from any other region.
For example, the energy resolution for K-shell events in the entire \emph{annular} volume is 1.8\%, and the distribution of phonon/ionization signal ratio for \emph{lateral} events is a Gaussian centered at 1 with a $\sigma$ of $1\%$.
These signals must come in a very large part from ballistic phonons with mean free paths much longer than the detector size that do not retain a memory of their point of origin.

Fig.~\ref{fig-ratios} also presents the distribution of K-shell events recorded at 15~V, normalized to the number of events at 66~V. 
The two plateaus above $r=1.05$ have similar heights and widths, although the 15~V plateau seems to taper off around $r=1.3$ instead of 1.4. 
This difference is small compared to the factor 3.83 increase of the NTL amplification between 15 and 66~V that should affect the $r$ distribution if NTL phonons were not contributing to the excess signal.

While the comparison of the $r$ distributions for the \emph{center}, \emph{intermediate}, \emph{annular} and \emph{lateral} selections provides information on the size of the lateral spread of the phonons giving rise to the excess, more insight on the depth of their point of origin can be obtained by studying the voltage-dependence of the $r$ distributions.

To quantify the evolution of the $r$ distribution as a function of the applied bias, the fractions of K-shell events with $r>1.2$ measured at absolute bias voltage values of 0, 8, 15, 30 and 66~V are shown in Fig.~\ref{fig-fraction}.
For the 66~V data, the values obtained at $-$66 and +66~V are plotted separately, and the value for the 1.3 keV L-shell events at $\pm$66~V is also shown.
All values at or above 30~V are consistent with an average fraction of 4.6\%, representing a rather large part (28\%) of all \emph{center} events.
This fraction thus does not depend on the event energy, or on whether the charges drifting toward the NbSi film are electrons or holes.
The small reduction to 3.5\% of the fraction of events with $r$ $>$ 1.2 at 8 and 15~V could reflect the onset of charge trapping and anisotropy effects at lower fields, as suggested by the evolution of the \emph{annular} volume fraction as a function of  bias in Table~\ref{tab-fraction}, and in particular the increase of the \emph{intermediate} volume at the expense of the \emph{center} one. 

More interesting is the observation that the fraction of events with $r$ $>$ 1.2 drops abruptly down to (0.5~$\pm$~0.1)\% at 0~V.
In the absence of the NTL effect, high-energy phonons can only come from the primary interaction.
In this case, the number of K-shell events with an excess in the $E_{inner}$ signal is proportional to the volume of the detector defined by the film surface and the depth corresponding to their average diffusion length.
In that hypothesis, the fraction of 0.5\% would correspond to a mean free path of the order of 1.2~mm.
This rough estimate is almost an order of magnitude larger than what would be expected~\cite{cdms-phonontime}, a possible indication that the production and transport of phonons in this region very close to the surface and to the NbSi film 
are  affected by stray fields and trapping centers. 

Despite these uncertainties, the fraction at 0~V clearly differs from the values above 3.4\% that are obtained as soon as bias of 8~V is applied, and the plateau is at 4.6\%. 
This is consistent with the excess $E_{inner}$ signal arising from the absorption of phonons emitted in the NTL process.
These are expected to propagate in a semidiffusive way over a relatively short distance~\cite{cdms-phonontime,phonon-wang,phonon-diffusion}. 
The phonons that are able to reach the inner film are those that are emitted within the last mm or $\mu$m distance travelled before reaching the detector surface covered by the film.
In a detector with a quasi planar field geometry, the emission of NTL phonons is spread evenly along the entire length of the field lines. 
The fraction of phonons that are emitted in the last portion of the field line is therefore independent of the depth of the event.
In the absence of a calibration of the efficiency and response of the NbSi film for phonons more energetic than the ballistic ones that contribute to the bulk of the signal, the size of the excess $E_{inner}$ signals provides no information on the actual diffusion length of the phonons.
However, the data of Fig.~\ref{fig-fraction} suggest that the effective mean free path of NTL phonons does not vary much between 8~V and 66~V, and is similar for those emitted by electrons and holes.

In this scenario, primary NTL phonons emitted far away from the film will not be detected.
The upper limit of this detection range is constrained to be small compared to the NbSi sensor dimension, the strongest experimental constraint being the width of 3~mm of the outer film.

\section{Rejection performance}

The constant fraction of events with $r$ $>$ 1.2 observed above 30~V, independently of signal energy and the polarity of the field, and their localization in the \emph{center} volume implies that the efficiency of this selection is 4.6\% when considering electron recoil events uniformly distributed in the entire detector volume.
In addition, the low efficiency of the $r$ $>$ 1.2 cut when applied to 0~V data suggests that this selection would disfavor non-ionizing events producing no NTL phonons, and could provide an efficient way to reject HO events. 
It could also significantly reduce the population of \emph{lateral} events, characterized by a poor charge collection.
The small efficiency of 4.6\% for evenly distributed electron recoil events could thus be compensated by a reduction of backgrounds caused by HO events and events with incomplete NTL amplification.

Selecting events with $r$ $>$ 1.2 means that $E_{inner}$ cannot provide a reliable measurement of the phonon energy, but this can be provided instead by the $E_{outer}$ signal, once it is corrected for the 95\% bias discussed in the previous section.
Using only one film to measure the phonon energy would however degrade the baseline phonon energy resolution from 103 eV to 138 eV.

\begin{figure}[tbh]
\begin{center}
\includegraphics[width=0.99\linewidth,trim= 0.2in 0in 0.5in 0in]{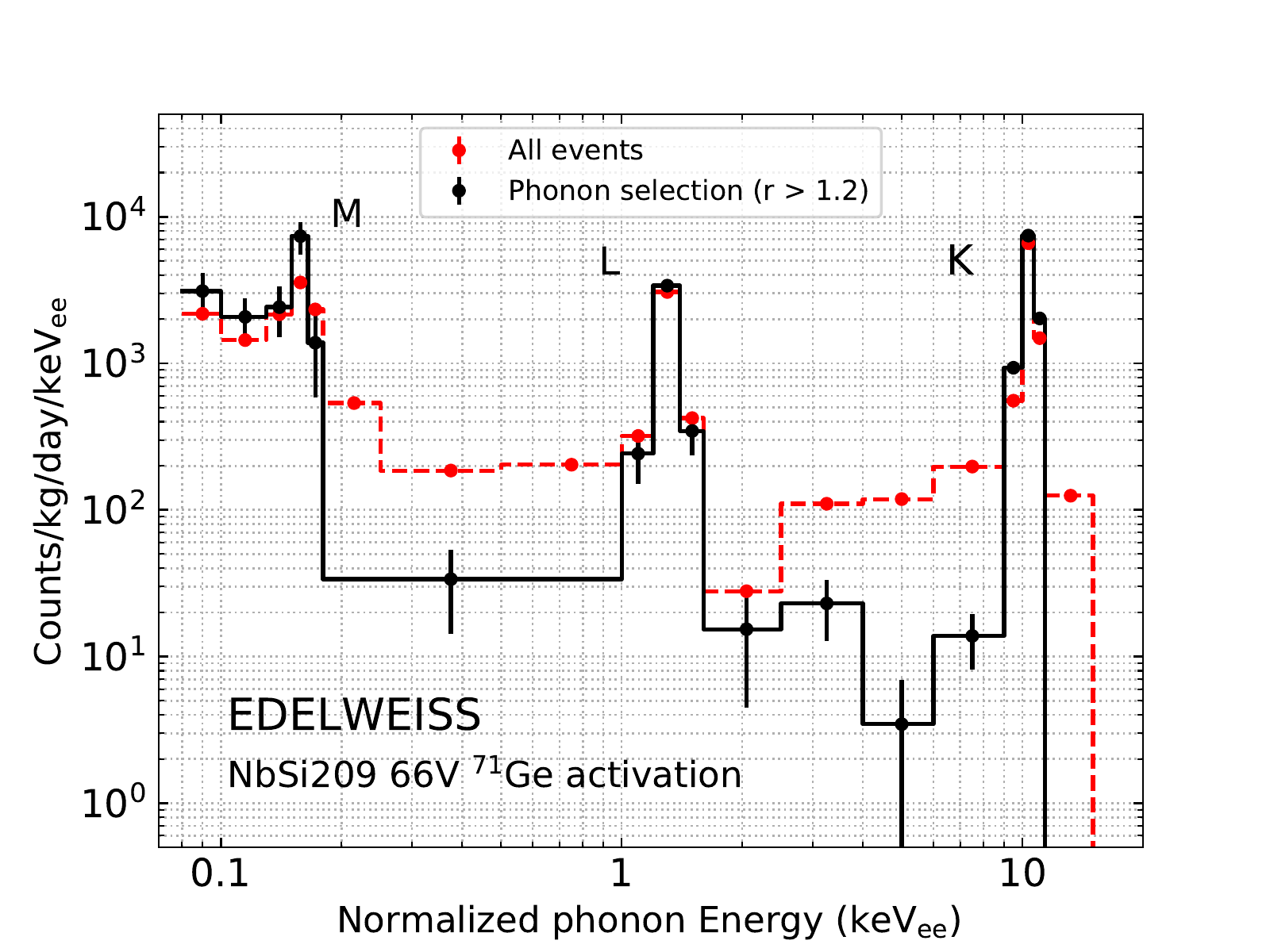}
\caption{Efficiency-corrected distributions of the normalized phonon energy (in keV$_{ee}$) collected at 66~V following the $^{71}$Ge activation of the crystal. Red: all events; black: events selected by the phonon cut ($r$ $>$ 1.2).  
}\label{fig-klm-reduction}
\end{center}
\end{figure}

\begin{figure}[tbh]
\begin{center}
\includegraphics[width=0.99\linewidth,trim= 0.2in 0in 0.5in 0in]{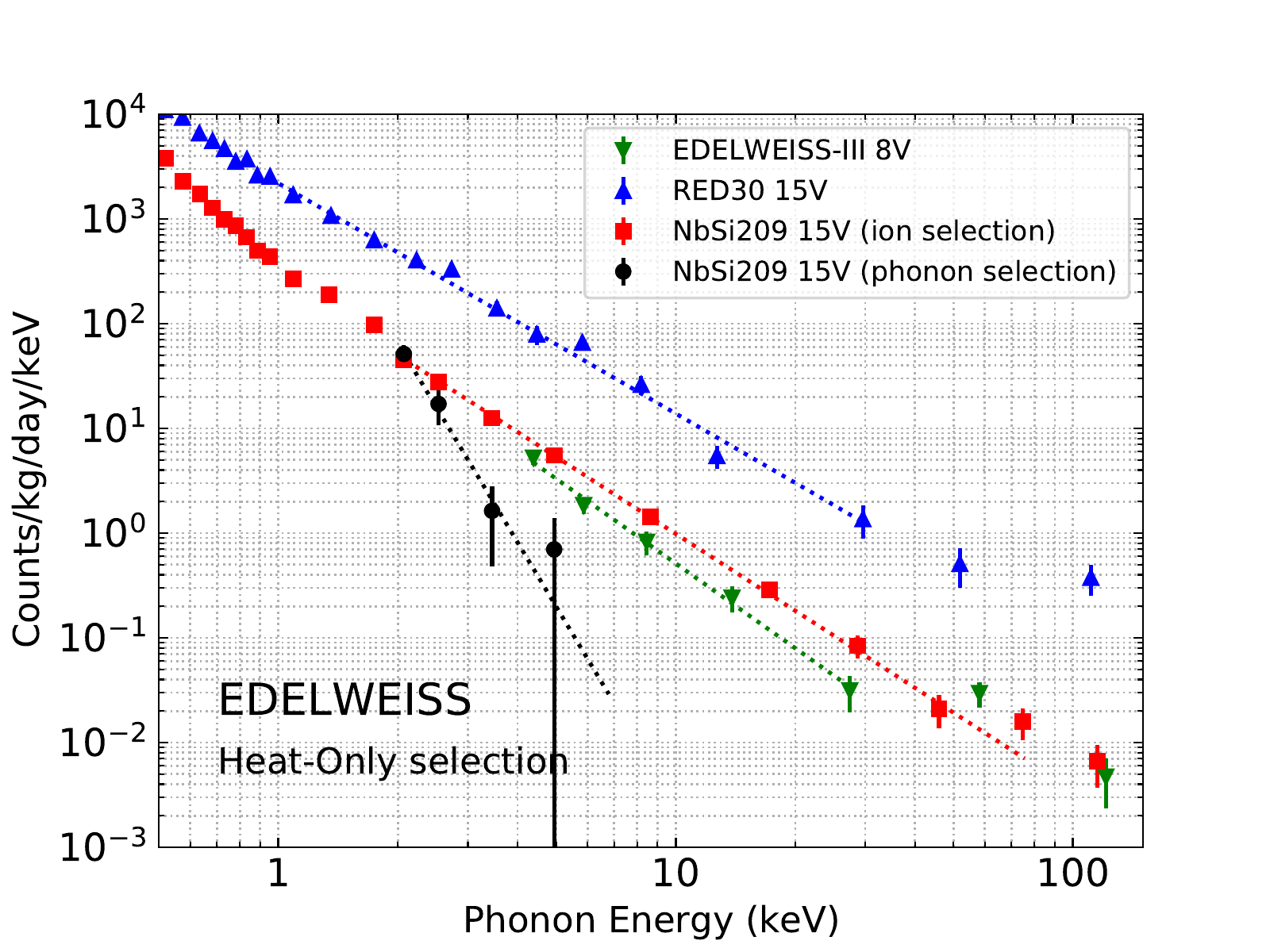}
\caption{Efficiency-corrected spectra of Heat-Only events selected by removing events with positive ionization signals. Green: EDELWEISS-III 0.82~kg detector; blue: the 33.4~g RED30 detector; and red: the NbSi209 0.20~kg detector. The efficiency-corrected spectra obtained with the NbSi209 detector by using the phonon cut ($r>1.2$) instead is shown in black. The dotted lines are power-law fits to guide the eyes.}\label{fig-ho}
\end{center}
\end{figure}

The advantage of applying the phonon cut to remove \emph{lateral} volume events is illustrated in Fig.~\ref{fig-klm-reduction}, showing the effect of that cut on the spectrum of events collected at 66~V in the month following the $^{71}$Ge activation of the detector. 
The efficiency correction takes into account the 4.6\% contribution of the $r$ $>$ 1.2 cut. 
The intensity of the 10.37 keV peak is recovered, as expected since the cut efficiency has been determined with K-shell events. 
More interestingly, the intensity of the L-shell peak is also recovered. 
The peak-over-background ratios of these two peaks are strongly enhanced because a large part of the inter-peak events are due to the poor charge collection of \emph{lateral}-volume events.
At low energy, the finite resolution on $r$ will result in a fraction of true $r=1$ events leaking through the $r>1.2$ cut.
Given the inner and outer films' baseline resolutions for the data in Fig.~\ref{fig-klm-reduction} (163 and 141 eV, respectively, for the weighted average of the $-$66~V and +66~V values in Table~\ref{tab-dataset}), the leakage fraction should exceed the value of the signal efficiency (4.6\%) at $E_{phonon}$~=~1.8~keV, corresponding to 0.08 keV$_{ee}$.

At low energy, the measurement of $r$ could be affected by uncertainties on the relative calibration of the inner and outer channel signals, and trigger biases.
The position of the M-shell peak in the two films indicates a possible 10\% shift at this low energy, that could push up $r$ by as much. 
Such a shift would push the energy where the background leakage fraction exceeds the signal cut efficiency to the M-shell peak, and explain the apparent absence of background reduction between 0.08 and 0.16 keV$_{ee}$ in Fig.~\ref{fig-klm-reduction}.

Increasing the cut on $r$ to values above 1.2 gives consistent results, but at the cost of a reduction of statistics. 
The use of values below 1.2 increases the leakage of the HO population at low energy.

To better evaluate the capacity of the $r>1.2$ cut to remove the HO background, it has been applied to a large-statistic sample of events recorded at $\pm$15~V with $E_{IonBottom}<0$.
Reducing the voltage bias from 66~V to 15~V improves the performance of the ionization cut at low energy and improves the purity of the selected HO sample.
The efficiency-corrected spectrum is shown in red in Fig.~\ref{fig-ho}.
The spectrum decreases as a power law of the energy between 2 and 100 keV, as observed in other EDELWEISS detectors~\cite{red30,edwtech} operated in the same cool-down.
At any given energy, the athermal phonon sensor measurement yields a lower number of HO events per unit mass than the Ge-NTD measurement of the 33.4~g RED30 detector~\cite{red30}, but comparable to that observed in a 0.82~kg EDELWEISS-III detector.
The use of an athermal phonon sensor does not eliminate by itself the problem of HO events.
However the situation is different when the $r>1.2$ cut is applied.
Despite the reduced statistics, the spectrum above 3 keV is reduced by a significant factor.
Only 2 events above 3 keV survive the cut out of the initial 594, which, given the efficiency of the $r>1.2$ cut, corresponds to a reduction factor greater than 5 at 90\% C.L..
Phonons produced in HO events are thus systematically detected by both films equally. 
Unequal detection yields among the two films is rare because it requires that the HO event originates at close proximity of one of them.
The situation is very similar to the reduction of the yield of K-shell events with $r>1.2$ when the bias is reduced to 0~V, as this turns off the emission of NTL phonons along the section of the field lines very close to the films.

As in Fig.~\ref{fig-klm-reduction}, the cut performance degrades rapidly at low energy due to the phonon baseline resolution (Table~\ref{tab-dataset}), and is no longer useful below 3 keV. 

\section{Impact on a Dark Matter Search}

The present limitation of the method to phonon energies above 3 keV precludes its use for the search of sub-GeV DM particles that involve nuclear recoils with kinetic energies less than 100 eV.
This type of search requires improved sensitivity for the NbSi TES, with sharper transitions and enhanced phonon collection.
The present performance proved to be sufficient for the above characterization of their response to localized ionization events and HO events.
It may also be relevant for Migdal searches involving electron recoils from the $n=3$ shell in Ge~\cite{nbsi-migdal}.
The energy spectrum of these electrons extends up to 200 eV$_{ee}$ and reaches the range where a tag is viable given the resolution of the NbSi209 prototype when operated at 66~V.

\begin{figure}[tbh]
\begin{center}
\includegraphics[width=0.99\linewidth,trim= 0.1in 0in 0.6in 0in]{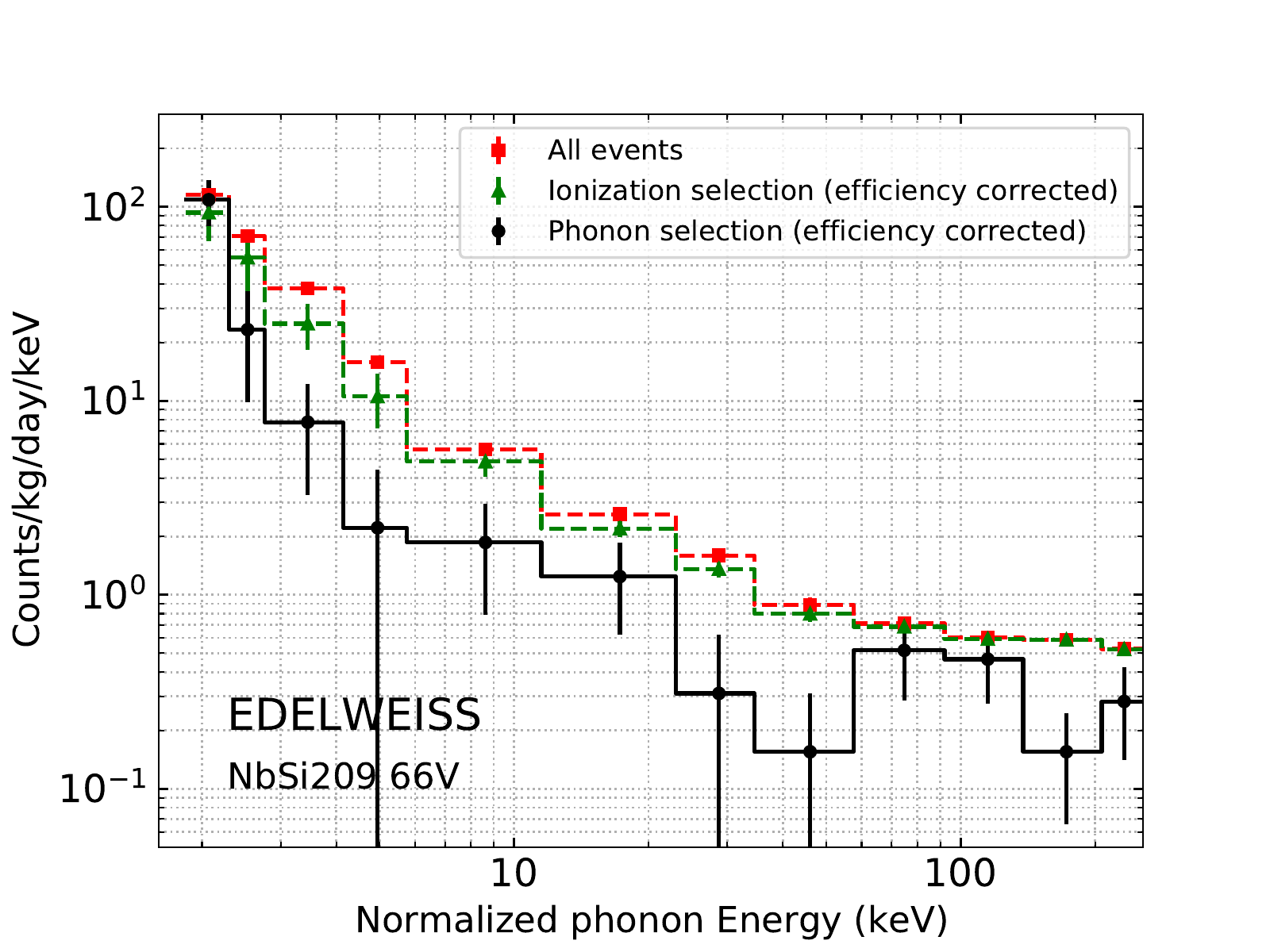}
\caption{Phonon energy distribution of the DM search data at $\pm$66~V. The plotted energy range (1.8--250 keV) corresponds to (0.08--10.9~keV$_{ee}$). Red squares: all events; green triangles: efficiency-corrected spectrum obtained with a 2$\sigma$ cut on the ionization; black circles: efficiency-corrected spectrum obtained with the phonon cuts.}\label{fig-searchdata}
\end{center}
\end{figure}

Fig.~\ref{fig-searchdata} compares the efficiency-corrected spectrum of the DM search data at $\pm$66~V with those obtained when electron recoils are specifically selected using either a cut on the ionization signal or on $r$.
Comparing with the HO spectrum for the 15~V data sample in Fig.~\ref{fig-ho}, that background should start to dominate the spectrum without the cuts of Fig.~\ref{fig-searchdata} around 4 keV, with a rate of 10 counts/kg/day/keV.
The cut used to produce the green dashed histogram in Fig.~\ref{fig-searchdata} is $E_{IonBottom}$ $>$ 2$\sigma_{IonBottom}$.
It only manages to produce a slight reduction of the rate between 3 and 6 keV.
This is due to the fact that the resolution of $\sigma_{IonBottom}$ = 325~eV$_{ee}$ for this data sample (see Table~\ref{tab-dataset}) means that the cut accepts 30\% of the HO population to leak at 4 keV (170 eV$_{ee}$).

The spectrum obtained with a cut on $r>1.2$ exhibits a significantly lower rate above 3 keV. 
This reduction can be attributed in part to suppression of some of the HO population. 
However, the phonon cut manages to reduce the rate by half at energies above  50 keV. The ionization cut fails to achieve a significant reduction in that range, where leakage effects should be negligible. 
This could indicate that the electron recoils selected by the phonon cut come from the \emph{center} region, and therefore benefit from self-shielding effects and a reduced exposure to surface events.

\begin{figure}[tbh]
\begin{center}
\includegraphics[width=0.99\linewidth]{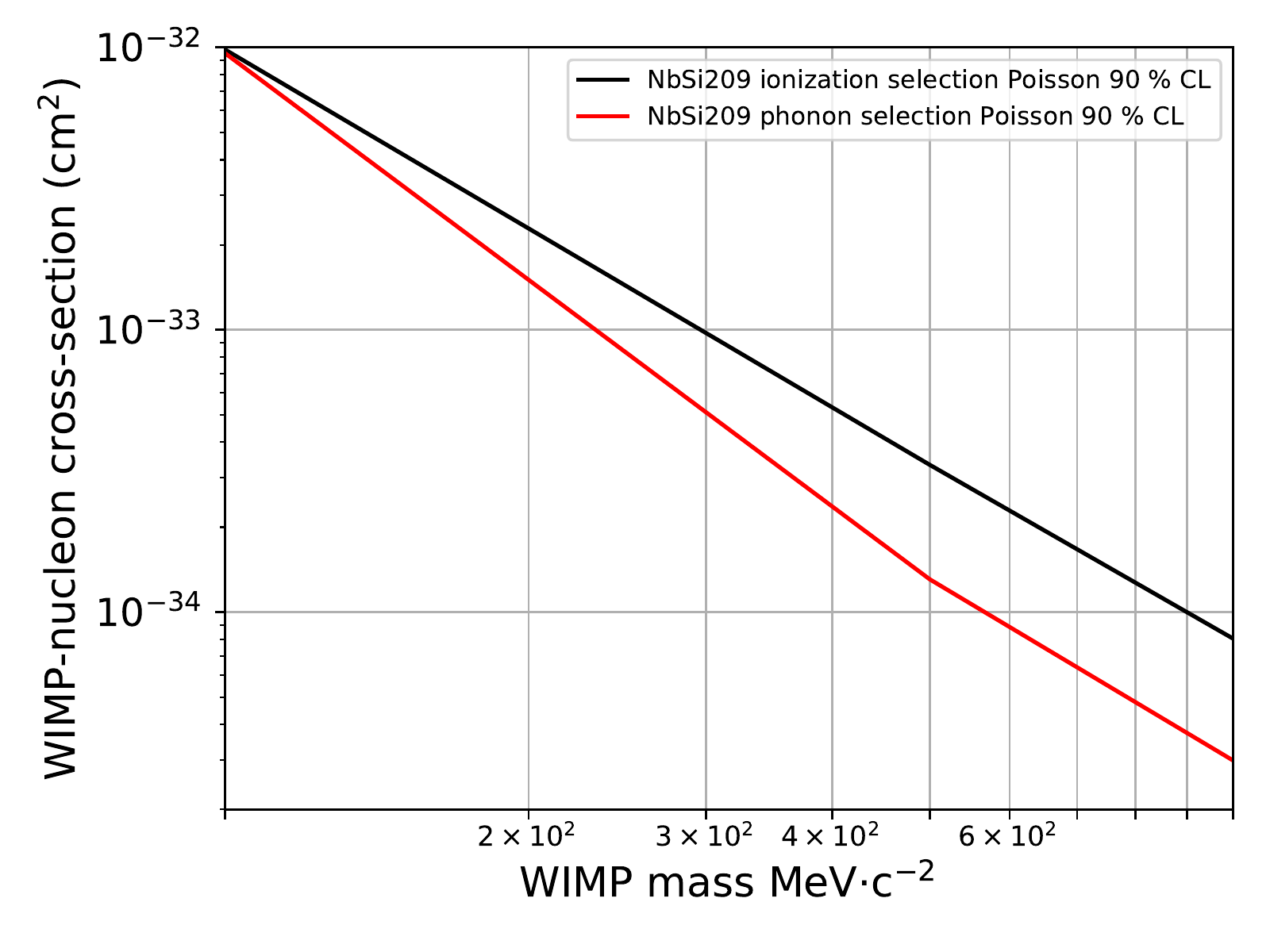}
\caption{90~\%C.L. limits on spin-independent cross sections for WIMP-nucleon interactions from Ref.~\cite{nbsi-migdal} obtained using an electron recoil selection based on the ionization signal (black) compared with those obtained in this work (red) using a phonon-based selection instead.}\label{fig-limits}
\end{center}
\end{figure}

The data displayed in Fig.~\ref{fig-searchdata} have been used to perform a search for DM particle interactions via the Migdal effect using the same analysis and starting with the same data set as in Ref.~\cite{nbsi-migdal}. 
The two searches differ by the cut used to reduce the background of HO events.
In Ref.~\cite{nbsi-migdal}, the cut is based on the ionization signal, and requires it to be consistent with full charge collection. 
Here, the ionization signal is not used at all, and the selection is only based on the ratio of the inner and outer phonon signals ($r>1.2$).
The 90\% C.L. limits on the cross-section for spin-independent WIMP-nucleon interaction cross-sections derived in both analyses are shown in Fig.~\ref{fig-limits}.
Both curves do not include Earth-shielding effects.
Their calculation in Ref.~\cite{nbsi-migdal} indicates that these are small in the mass and cross-section regime and should not affect the relative comparisons.
The limits obtained in both cases are in a domain excluded by other experiments~\cite{cedex,xenon-migdal}, but their comparison helps in understanding the potential of an athermal phonon-based tag in the context of DM searches. 
The limits are identical for a WIMP particle with a mass 100 MeV/c$^2$.
This can be expected as the optimal region of interest is in the 2--3 keV range in this case, an energy at which the $r>1.2$ is no longer helpful. 
For a mass of 1 GeV$/c^2$, the use of the phonon cut improves the exclusion limit by a factor 2.8.
This can be explained by the optimal region of interest being in the range from 2.9 to 4.9 keV, for which the phonon cut is effective at reducing the background, as observed in Fig.~\ref{fig-searchdata}.

A phonon-based cut can thus be used to reduce backgrounds from HO and surface events.
However, in order to be truly useful for sub-GeV DM particle searches, the energy resolution of the present device must be improved by two orders of magnitude.

\section{Anticipated improvements}

Work is ongoing to remedy the two problems of the limited resolution of the present TES, as well as the small fiducial mass corresponding to the applied phonon cut.
A better control of the NbSi film uniformity would improve the sharpness of the transition and therefore the energy resolution.
A more original solution to remove the HO population would be to optimize the NbSi film to single-electron level sensitivity to provide a tag for ionizing events, and to rely on the established ~eV resolution obtained by NTL-boosted Ge-NTD for the energy measurement~\cite{red30}. 
Such an improvement in film sensitivity can be achieved by reducing it to a single short $\sim$mm strip, inspired by the design of Superconducting Single Electron Device (SSED), and by operating it well below its critical temperature $T_c$ such that only a massive phonon shower can trigger a transition.
To increase the fiducial volume while keeping a clean separation between the diffuse phonon production of HO events from a spatially concentrated phonon production from the NTL amplification, the  asymmetric field configuration provided by a point-contact electrode will be used: the small SSED will be used as one polarity electrode, while most of the rest of the detector surface will be covered by an enveloping electrode.
In that way, the emission of phonons from the NTL amplification of ionizing events from the bulk of the volume will occur in the small high-field region very close to the SSED. 
At the same time, the small SSED size reduces its efficiency to detect uniformly distributed primary phonons.
In this way, the increase of the fiducial volume is not made at the expense of losing the ability to reject populations that are not accompanied by charges drifting through this high-field region, such as HO events and events where most of the charge is trapped on the surfaces. 

\section{Conclusion}

In this paper, we described a technique able to tag ionizing events using a pair of NbSi TES athermal phonon sensors. 
The difference between the signals in the two films is used to identify the highly localized primary phonons generated by the Neganov-Trofimov-Luke effect associated with the drift of the electrons and the holes as they approach one film or the other.
This tag selects ionizing events occurring in regions traversed by field lines that intersect one of the two TES.
The use of this tag is demonstrated to reduce the population of Heat-Only (HO) events, and of electron recoil events originating from the outer perimeter of the detector.
Notably, the HO population is reduced by more than a factor 5 for phonon energies above 3 keV.
The performance is limited by the finite energy resolution of the current TES prototype.
This selection method is nevertheless able to improve by a factor 2.8 some of the Dark Matter limits obtained in Ref.~\cite{nbsi-migdal} using the same data set.
Establishing that phonons originating from the Neganov-Trofimov-Luke amplification process can be localized using an NbSi TES leads the way for further major improvements of the technique for sub-GeV Dark Matter searches. 

\begin{acknowledgments}
The help of the technical staff of the Laboratoire Souterrain de Modane and the participant laboratories is gratefully acknowledged. 
The EDELWEISS project is supported in part by the French Agence Nationale pour la Recherche (ANR-21-CE31-0004), by the P2IO LabEx (ANR-10-LABX-0038) in the framework ``Investissements d'Avenir'' (ANR-11-IDEX-0003-01), and the LabEx Lyon Institute of Origins (ANR-10-LABX-0066) within the frameork of the program France 2030, also operated by the National Research Agency of France. 
B.J. Kavanagh thanks the Spanish Agencia Estatal de Investigaci\'on (AEI, MICIU) for the support to the Unidad de Excelencia Mar\'ia de Maeztu Instituto de F\'isica de Cantabria, ref. MDM-2017-0765.
BJK also acknowledges funding from the Ramón y Cajal Grant RYC2021-034757-I, financed by MCIN/AEI/10.13039/501100011033 and by the European Union “NextGenerationEU”/PRTR.
\end{acknowledgments}

\newcommand{\arXiv}[1]{\href{https://arxiv.org/abs/#1}{arXiv:#1}}
\newcommand{\oldarXiv}[1]{\href{https://arxiv.org/abs/#1}{#1}}
\newcommand{\DOI}{https://doi.org}

\end{document}